\documentstyle[aps,psfig,epsfig]{revtex}

\newcommand{\pr}{\prime}
\newcommand{\na}{\nabla}

\newcommand{\ep}{\epsilon}

\newcommand{\vphi}{\varphi}

\newcommand{\lraw}{\longrightarrow}

\newcommand{\Gafunc}[1]{\Gamma(#1-\frac{D}{2})}
\newcommand{\quadiv}[1]{\frac{#1 N_c}{(4\pi)^{D/2}}
       (\frac{\mu^2}{m^2})^{\epsilon/2}\Gafunc{1}}
\newcommand{\logdiv}[1]{\frac{#1 N_c}{(4\pi)^{D/2}}
       (\frac{\mu^2}{m^2})^{\epsilon/2}\Gafunc{2}}

\newcommand{\pa}{\partial}

\begin{document}

\draft
\title{Chiral Expansion at Energy Scale of $\rho$-Mass}
\author{Xiao-Jun Wang\footnote{E-mail address: wangxj@mail.ustc.edu.cn}}
\address{Center for Fundamental Physics,
University of Science and Technology of China\\
Hefei, Anhui 230026, P.R. China}
\author{Mu-Lin Yan\footnote{E-mail address: mlyan@staff.ustc.edu.cn}}
\address{CCST(World Lad), P.O. Box 8730, Beijing, 100080, P.R. China \\
  and\\
 Center for Fundamental Physics,
University of Science and Technology of China\\
Hefei, Anhui 230026, P.R. China\footnote{mail
address}}
\date{\today}
\maketitle

\begin{abstract}
{We study chiral expansion at $m_\rho$-scale in framework of
chiral constituent quark model. The lowest vector meson resonsances
are treated as composited fields of constituent quarks. We illustrate
that, at energy scale of $\rho$-meson mass, the chiral expansion expansion
converges slowly. Therefore, it is possible to construct a well-defined
chiral effective field theory at this energy scale, but high order
correction of chiral expansion must be included simultanously. The
one-loop correction of pseudoscalar mesons is also studied systematically.
The unitarity of the model is examined and Breit-Wigner formula for
$\rho$-meson is obtained. The prediction on on-shell 
$\rho\rightarrow\pi\pi$ and $\rho\rightarrow e^+e^-$ decays agree with 
data very well.}
\end{abstract}

\section{Introduction}

Although at very low energy the chiral perturbative 
theory(ChPT)\cite{GL85a} provides an excellent description on interaction
of pseudoscalar mesons as well as perturbative QCD works in high energy,
it has to be recognized that, so far, we only have a little knowledges
concerning underlying dynamical detail at low energy. In particular, in
energy region between ChPT($\mu\sim 500$MeV) and chiral symmetry
spontaneously broken(CSSB) scale($\Lambda_{\rm CSSB}\sim 1.2$GeV), we do
not know how to perform rigorous calculation based on underlying dynamics
or symmetry at all, since those well-defined QCD expansion($\alpha_s$
expansion, low energy expansion, etc.) converge slowly or even diverge
here. Therefore, some phenomenological models(e.g., hidden local symmetry
model\cite{Bando}, antisymmetry tensor model\cite{Ecker89}, WCCWZ
model\cite{Brise97}) are constructed for capturing the physics
in this energy region. All of these models base on chiral symmetry
and only include the lowest order coupling between vector meson
resonances and pseudoscalar mesons, i.e., all couplings are
momentum-independent and those coupling constants are determined
by experiment instead of underlying dynamics. It is obvious that
these models only provide very rough physical picture on the lowest
vector meson resonances. In general, if a well-defined effective
field theory indeed exists at this energy scale, all couplings
should be momentum-dependent instead of constants, or these
couplings can yield convergence momentum expansion. Since vector
meson masses are much larger than pseudoscalar mesons, it is
different from ChPT that high order correction of the chiral expansion 
plays important role here. The purpose of this paper is to provide
a possible method to systematically study the chiral expansion
at $\rho$-mass scale.

Another fact inspires us to perform this research that the energy
scale of CSSB is larger than masses of the lowest vector meson
resonances. Therefore, if the chiral expansion is in
powers of $p^2/\Lambda_{CSSB}^2(p\sim m_\rho)$, it converges slowly.
Obviously, in this case contribution from high order terms of momentum
expansion plays an important role. However,
since we do not know how to derive a low energy effective field theory
from QCD directly, we have to construct the low energy effective theory
in terms of some approaches with features of low energy QCD. In general,
there are two different approaches for studying dynamics of light-flavor
$0^-$ and $1^-$ mesons in framework of effective field theory: One is
the method of ChPT, which only needs to assume a certain realization of
chiral symmetry for vector mesons(it is standard for symmetry
realization of pseudoscalar mesons). The effective lagrangian is written
based on these symmetrical reqiurement and is expanded in powers of
external momentum. The hidden local symmetry model et.
al.\cite{Bando,Ecker89,Bijnen97} belong to this approach. A recent review
is in\cite{Brise97}. In principle, this method can be treated as
approximate symmetry pattern of QCD and it's symmetry spontaneously
breaking if symmetrical realization of vector mesons is right.
Unfortunately, the method of ChPT is impractical here since our
calculation in this paper must go beyond the lowest order, and
number of free parameters will increase very rapidly. Another
approach is method of chiral quark model(ChQM). The
Manohar-Georgi(MG) model\cite{GM84} and Extend
Nambu-Jona-Lasinio(ENJL)\cite{NJL,Bijnen96a,Hatsuda94} model et. al.
all belong to this approach. In chiral quark model, the vector
meson fields are coupling to quark fields. There are no kinetic
terms for vector mesons. Therefore they are treated as composited fields
of quarks instead of independent degrees of freedom. Effective lagrangian
describing dynamics of vector mesons is yielded via quark loop effects.
The ChQM approach has some advantages which are try to reflect some
underlying dynamical constrains and provide elegant description on some
features of low-energy QCD. The main advantage is that number of free
parameters does not increase with expansion order rising. So that it is
possible to investigate high order correction of momentum expansion
systematically. Of course, the ChQM approach is only treated as model
instead of rigorous theory, since due to lack of knowledge on QCD low
energy behavior, one has to add quark-meson coupling or more undelying
four fermion coupling handly according to requirement of symmetry.
Although this is a disadvantage of the ChQM approach, the ChQM and its
extension\cite{Wein79}-\cite{WY99} have been studied continually
during the last two decades and got great successes in different
aspects of phenomenological predictions in hadron physics. In this
paper, following spirit shown in MG model, we will construct the
chiral constituent quark model(ChCQM) for the lowest vector meson
resonances, and provide systematic investigation on dynamics of
on-shell $\rho\rightarrow\pi\pi$ and $\rho-e^+ e^-$ decays. 

The simplest version of ChQM which was originated by
Weinberg\cite{Wein79}, and developed by Manohar and
Georgi\cite{GM84} provides a QCD-inspired description on the
simple constituent quark model. In view of this model, in the
energy region between the CSSB scale and the confinement scale
($\Lambda_{QCD}\sim 0.1-0.3 GeV$), the dynamical field degrees of
freedom are constituent quarks(quasi-particle of quarks), gluons
and Goldstone bosons associated with CSSB(it is well-known that
these Goldstone bosons correspond to lowest pseudoscalar octet).
In this quasiparticle description, the effective coupling between
gluon and quarks is small and the important interaction is the
coupling between quarks and Goldstone bosons. Simultaneously, this
simple model provide a very rough configuration about baryons,
which is baryons can be treated as bound states of three
constituent quarks. Thus, from baryon masses, masses of
constituent quarks are approximately estimated as 360MeV for u,d
flavor and 540MeV for s flavor\cite{GM84}. We notice that the
masses for u,d flavor are very close to $m_{\rho}/2$. In addition,
the binding energy of nucleons is expected to be large due to
stability of nucleons. It implies that true masses of constituent
quarks are larger than our above estimation even $m_{\rho}/2$.
Naturally, it is allowed to treat the lowest vector meson
resonances as composite fields of constituent quark and antiquarks
instead of independent dynamical degree of freedom. Thus, dynamics
of vector mesons will be generated by constituent quark loops.
This approach is foundation of our study in this paper.

Furthermore, we must point out that, role of pseudoscalar meson
fields in constituent quark model is different from other two kinds
of ChQM: ENJL model and chiral current quark model\cite{Li95a,WY99}. In
the latter the pseudoscalar mesons are composited fields of
current quark and antiquarks. So that they are not independent
dynamical field degree of freedom of these models and dynamics of
$0^-$ mesons is generated by current quark loops. However, in the
former the pseudoscalar mesons are independent dynamics degree of
freedom instead of bound states of constituent quarks. This implies
that the model is "part renormalizatable"(for kinetic term of $0^-$
mesons). There is no so called double counting problem in this
model due to the following two reasons: 1) The constituent quark
fields and pseudoscalar mesons(as Goldstone bosons) are generated
simultaneously by CSSB. 2) Phenomenologically, the masses of
constituent quarks are much larger than of $0^-$ mesons. Therefore,
if pseudoscalar fields are bound states of constituent quarks, such
large binding energy is anti-intuitive. Although ENJL model or
chiral current model provide a possibility to probe underly
dynamics structure on pseudoscalar mesons, they are failed to touch the
goal of this paper, since these models can not yield a sufficient
large paramter consistently to make the chiral expansion at energy scale
$\mu\sim m_\rho$ be convergent.

It has been known that, there are some rather different forms on low
energy effective lagrangian including spin-1 meson resonances, and the
different types of couplings contained in them. Every approach corresponds
to a different chioce of fields for the spin-1 mesons, and they are in
principle equivalent. From the viewpoint of chiral symmetry only, an
alternative scheme for incorporating spin-1 mesons was suggested by
Weinberg\cite{Wein68} and developed further by Callan, Coleman et.
al\cite{CCWZ69}. In this treatment, all spin-1 meson resonances transform
homogeneously under a non-linear realizations of chiral $SU(3)$, which are
uniquely determined by the known transformation properties under
the vectorial subgroup $SU(3)_V$(octets and singlet). This is an
attractive symmetry property on meson resonances and quite nature
in ChCQM, since in this model all dynamical field degree of freedom
are associated with CSSB so that lagrangian is explicitly invariant
under local $SU(3)_V$ transformation. Of course, it is not
necessary to describe the degrees of freedom of vector and
axial-vector mesons by antisymmetric tensor fields\cite{Ecker89},
and there are other phenomenological successful attempts to
introduce spin-1 meson resonances as massive Yang-Mills
fields\cite{Bando,Li95a,Meissner88}. We will show that, in ChCQM,
vector representation for vector mesons is nature and more
convenient than tensor representation.

In past thirties years, various approachs have been attempted to
predict hadron phenomenology. Many of these approachs on vector
mesons are motivated by phenomenologically successful ideas of
vector-meson dominance(VMD) and universal coupling\cite{Sakurai69,AFFR73}. 
In ChCQM, we only need start from bound states approachs for
vector meson resonances and transformation properties of their vector
repsentation. The VMD and universal coupling will be naturally predicted
by the model in stead of input. Consequently, other phenomenological
relation, such as KSRF sum rules are yielded too. This is
anthor advantage of ChCQM. It should be noted that, in making comparisons
between ChCQM and other approachs, we need carefully distinguish
features coming from the choice of field from those coming from
phenomenological requirments. The former are not physical, controlling
merely the off-shell behaviour of scattering amplitudes. The later do have
physical consequences, such as relations between on-shell amplitudes for
different processes. Furthermore, for purpose of this paper, a nature
problem appears: Can phenomenological results obtained in leading order
still be kept when high order of momentum expansion are considered? This
problem will be carelly study in this paper.

It can be known from ChPT, if an effective field theory is constructed in
powers of momentum expansion, loop graphs of this field theory which comes
from lower order terms will contribute to higher order terms. A nature
agruement is loop graphs of effective field theory of QCD are suppressed
by $1/N_c$ expansion\cite{tH74}. However, for physical value $N_c=3$, this
suppression is not suficiently small, so that we can not omit contribution
from hadron loop graphs in our calculation(especially, one-loop graph).
Due to large mass gap between pseudoscalar mesons and vector meson, it is
reasonable assumption that dominant contribution comes from one-loop
graphs of pseduoscalar meson. The dynamics including one-loop contribution
is very different from one of leading order, for instance, imaginary of
${\cal T}$-matrix of this effective theory will be yielded. Naturally, it
is very difficult to deal with ultraviolet(UV) divergence from hadron
loops in a framework of non-renormalizable field theory. Fortunately,
loop effects of pseudoscalar meson cause $\phi-\omega$ mixing which will
destory OZI rule if this contribution is large. Thus we can cancel all UV
divergence from $0^-$ meson loops in terms of OZI rule.

It is different from some approachs that\cite{Ecker89,Bijnen96a,Li95a},
according to proposition of this paper, physics about axial-vector
meson resonances, $a_1$(1260), can not be studied here. Since
the chiral expansion in this energy region is not convergent. In
fact, this problem exists in all approachs including axial-vector meson
resonances. Of course, from alternative viewpoint, those approachs can be
understood and only be understood as phenomenological models in the
leading order. In this paper, since we will provide a rigorous treatment
on the chiral expansion, we only focus our attention on vector mesons.

The paper is organized as follows. In sect. 2 the chiral
constituent quark model with vector meson resonances are
constructed, and the effective lagrangian at leading order are
derived. This effective lagranguan is equivalent to WCCWZ lagrangian given
by Brise\cite{Brise97}. In sect. 3 we will calculate effective
vertices for $\rho-\gamma$ mixing, $\rho\rightarrow\pi\pi$, 
$\gamma\rightarrow\pi\pi$ and four-pseudoscalar meson coupling. Those
effective vertices is generated by constituent quark loops, and include
all orders correction of momentum expansion. In sect. 4, one-loop
correction generated by pseudoscalar mesons is calculated systematically.
Our goal is to estimate hadronic one-loop contribution in
$\rho\rightarrow e^+e^-$ and $\rho\rightarrow\pi\pi$ decay amplitude. The
Breit-Wigner formula for $\rho$-propagator is obtained. The unitarity of
this effective theory is also examined explicitly. The numerical result is
in sect. 5 and sect. 6 is devoted to summary.

\section{ChCQM with Vector Meson Resonances}
\setcounter{equation}{0}
\setcounter{figure}{0}

\subsection{Construction of ChCQM}
The QCD lagrangian with three flavour current quark fields
$\bar{\psi}=(\bar{u},\bar{d},\bar{s})$ is,
\begin{eqnarray}
\label{2.1}
&&{\cal L}_{QCD}(x)={\cal L}_{QCD}^{0}+{\cal L}_{\rm f},
     \nonumber \\
&&{\cal L}_{\rm f}=\bar{\psi}(\gamma \cdot v+\gamma
    \cdot a\gamma_5)\psi-\bar{\psi}(s-i\gamma_5p)\psi. \nonumber \\
\end{eqnarray}
For our purpose we only pay attention to ${\cal L}_{\rm f}$. Here
the fields $v_\mu, a_\mu$ and $p$ are $3\times3$ matrices in
flavour space and denote respectively vector, axial-vector and
pseudoscalar external fields. $s={\cal M}+s_{_{\rm external}}$,
where $s_{_{\rm external}}$ is scalar external fields and ${\cal
M}$=diag($m_u,m_d,m_s$) is current quark mass matrix with three
flavors.

The introduction of external fields $v_\mu$ and $a_\mu$ allows for
the global symmetry of the lagrangian to be invariant under local
$SU(3)_L \times SU(3)_R$, i.e., with $g_L,g_R\in SU(3)_L\times SU(3)_R$,
the explicit transformations of the different fields are
\begin{eqnarray}
\label{2.2}
      &&\psi(x)\rightarrow g_R(x)\frac{1}{2}(1+\gamma_5)\psi(x)
           +g_L(x)\frac{1}{2}(1-\gamma_5)\psi(x), \nonumber \\
      &&l_\mu\equiv v_\mu-a_\mu \rightarrow g_L(x)l_\mu g_L^{\dagger}(x)
           +ig_L(x)\partial_\mu g_L^{\dagger}(x), \nonumber \\
      &&r_\mu\equiv v_\mu+a_\mu \rightarrow g_R(x)r_\mu g_R^{\dagger}(x)
           +ig_R(x)\partial_\mu g_R^{\dagger}(x), \nonumber \\
      &&s+ip \rightarrow g_R(x)(s+ip)g_L^{\dag}(x).
\end{eqnarray}

Now let energy descend until chiral symmetry is spontanoeusly
broken. Below this energy scale, the coupling becomes
strong and perturbative QCD can no longer be done, so that we need
some effective models(quark model, pole model, Skyrme model...) to
approach low energy behaviours of QCD. A successful attempt is
achieved by non-linear realization of spontanoeusly broken global
chiral symmetry introduced by Weinberg\cite{Wein68}. This
realization is obtained by specifying the action of global chiral
group $G=SU(3)_L\times SU(3)_R$ on element $\xi(\Phi)$ of the coset
space $G/SU(3)_{_V}$:
\begin{equation}\label{2.3}
\xi(\Phi)\rightarrow
g_R\xi(\Phi)h^{\dag}(\Phi)=h(\Phi)\xi(\Phi)g_L^{\dag},\hspace{1in}
 h(\Phi)\in H=SU(3)_{_V}.
\end{equation}
Explicit form of $\xi(\Phi)$ is usually taken
\begin{equation}\label{2.4}
\xi(\Phi)=\exp{\{i\lambda^a \Phi^a(x)/2\}},
\end{equation}
where the Goldstone boson $\Phi^a$ are treated as pseudoscalar
meson octet. The compensating $SU(3)_{_V}$ transformation $h(\Phi)$
defined by Eq.(~\ref{2.3}) is the wanted ingredent for a non-linear
realization of G. In practice, we shall be interested in transformations
of constituent quark fields and spin-1 meson resonances under
$SU(3)_{_V}$. The constituent quarks $\bar{q}=(\bar{q}_u,
\bar{q}_d, \bar{q}_s)$ are defined as fields whose quantum numbers are
same as current quarks $\bar{\psi}$. The $q,\bar{q}$ transform as
matter fields of $SU(3)_{_V}$:
\begin{equation}\label{2.5}
  q\lraw h(\Phi)q, \hspace{1in} \bar{q}\lraw \bar{q}h^{\dag}(\Phi).
\end{equation}
The spin-1 meson resonances transform homogeneously as octets and singlets
under $SU(3)_V$. Denoting the multiplets generically be $O_\mu$(octet)
and $O_{1\mu}$(singlet), the non-linear realization of G is given
by
\begin{equation}\label{2.6}
  {\cal O}_\mu\rightarrow h(\Phi){\cal O}_{\mu}h^{\dag}(\Phi),
   \hspace{1in}{\cal O}_{1\mu}\rightarrow {\cal O}_{1\mu}.
\end{equation}
More convenience, due to OZI rule, the vector and axial-vector octets and
singlets are combined into a single ``nonet'' matrix
$$ N_\mu={\cal O}_\mu+\frac{I}{\sqrt{3}}{\cal O}_{1\mu},\hspace{0.8in}
N_\mu=V_\mu,A_\mu, $$
where
\begin{equation}
\label{2.7}
   V_\mu(x)={\bf \lambda \cdot V}_\mu =\sqrt{2}
\left(\begin{array}{ccc}
       \frac{\rho^0_\mu}{\sqrt{2}}+\frac{\omega_\mu}{\sqrt{2}}
            &\rho^+_\mu &K^{*+}_\mu   \\
    \rho^-_\mu&-\frac{\rho^0_\mu}{\sqrt{2}}+\frac{\omega_\mu}{\sqrt{2}}
            &K^{*0}_\mu   \\
       K^{*-}_\mu&\bar{K}^{*0}_\mu&\phi_\mu
       \end{array} \right).
\end{equation}
As momentioned in Introduction, in this formalism we can not study
physics at axial-vector meson mass scale consistently, but there is
no problem when axial-vector mesons appear as off-shell fields.
Thus axial-vector meson resonances $A_\mu$ will affect low
energy dynamics of pseudoscalar fields through $A_\mu-\pa_\mu\Phi$ mixing.
Therefore we still remain fields $A_\mu$ here. But it should be remembered
that $A_\mu$ only appear as intermediate states, and in this paper we will
remove them after we diagonize $A_\mu-\pa_\mu\Phi$ mixing.

Due to introduction of external fields $v_\mu$ and $a_\mu$, the
model can be extended to be invariant under $G_{\rm glocal}\times
G_{\rm local}$. So that it is convenient to put pseudoscalar fields
and external vector and axial-vector fields in $SU(3)_{_V}$
invariant field gradients
\begin{equation}\label{2.8}
  \Delta_\mu=\frac{1}{2}\{\xi^{\dag}(\pa_\mu-ir_\mu)\xi
          -\xi(\pa_\mu-il_\mu)\xi^{\dag}\},
\end{equation}
and connection
\begin{equation}\label{2.9}
  \Gamma_\mu=\frac{1}{2}\{\xi^{\dag}(\pa_\mu-ir_\mu)\xi
          +\xi(\pa_\mu-il_\mu)\xi^{\dag}\}.
\end{equation}
Under non-linear realization of chiral SU(3) $\Gamma_\mu$
transforms as follow:
\begin{equation}\label{2.10}
  \Gamma_\mu\lraw h\Gamma_\mu h^{\dag}+h\pa_\mu h^{\dag}.
\end{equation}
Without external fields, $\Gamma_\mu$ is the usual natural connection on
coset space. Since the above transformation is local we are led to define
a covariant derivative
\begin{equation}\label{2.11}
 d_\mu{\cal O}=\pa_\mu{\cal O}+[\Gamma_\mu,{\cal O}],
\end{equation}
ensuring the proper transformation
\begin{equation}\label{2.12}
d_\mu{\cal O}\lraw h(\Phi)d_\mu{\cal O}h^{\dag}(\Phi).
\end{equation}

In addition, when we want going beyond chiral limit, the current quark
mass enter dynamics by means of the following $SU(3)_{_V}$ invariant form
\begin{equation}\label{2.13}
\frac{1}{4B_0}(\xi\chi^{\dag}\xi+\xi^{\dag}\chi\xi^{\dag})
+\frac{1}{4B_0}\kappa(\xi\chi^{\dag}\xi
  -\xi^{\dag}\chi\xi^{\dag})\gamma_5,
\end{equation}
with $\chi=2B_0(s+ip)$.

By using on similar discussion, Manahor and Georgi provide a simple
pattern of ChCQM for understanding the physics between CSSB scale and
quark confinement scale\cite{GM84}. The MG model are described by the
following chiral constituent quark lagrangian
\begin{equation}\label{2.14}
{\cal L}_{\rm MG}=i\bar{q}\gamma\cdot(\pa+\Gamma+g_{_A}\Delta\gamma_5)q
    -m\bar{q}q+\frac{F^2}{16}<\nabla_\mu U\nabla^\mu U^{\dag}>,
\end{equation}
where $U(\Phi)=\xi^2(\Phi)$, $g_{_A}$ is coupling constant of axial-vector
current whose value $g_{_A}\simeq 0.75$ can be fitted by $n\rightarrow
pe^-\bar{\nu}_e$ decay. The $<...>$ denotes trace in SU(3) flavour space
and covariant derivative is defined as follows:
\begin{eqnarray}\label{2.15}
\nabla_\mu U&=&\pa_\mu U-ir_\mu U+iUl_\mu=2\xi\Delta_\mu\xi,
  \nonumber \\
\nabla_\mu U^{\dag}&=&\pa_\mu U^{\dag}-il_\mu U^{\dag}+iU^{\dag}r_\mu
  =-2\xi^{\dag}\Delta_\mu\xi^{\dag}.
\end{eqnarray}
In lagrangian(~\ref{2.14}), mass of constituent quarks $m$ is a paramter
relating to CSSB. Here we treat that mass difference of constituent quarks
for different flavors are caused by current quark masses. According to the
discussions presented in the Introduction, it is theoretically 
self-consistent when kinetic term of pseudoscalar mesons is introduced
initially. Thus MG model is renormalizable for kinetic term and the lowest
order interaction term of pseudoscalar meson. The high order interaction
for $0^-$ mesons will be generated by both of loop effects of quarks and
loop effects of the lowest order interaction of mesons.
By means of M-G model, the quark mass-independent low energy coupling
constants have been derived in Refs.\cite{WY98,Esp90}. It is remarkable
that the predictions of this simple model are in agreement with the
phenomenological values of $L_i$ in ChPT. This means the low energy limit
M-G model is compatible with ChPT in chiral limit. In the baryon physics,
the skyrmion calculations show also that the predictions from M-G model
are reasonable\cite{Chan,Ait85,LY92}.

This simple model provides an useful description on physics between CSSB
scale ($\mu\sim 1.2$GeV) and quark confinement scale ($\mu\sim
0.1-0.3$GeV). That is if we live in a world with this energy region only,
perhaps we will not think about what is quark confinement very much, or
even can not discover what QCD is at all. We can construct a consistent
``field theory'' in terms of Goldstone bosons and those ``fake element
particles''-constituent quarks. We will be perfectly satisfied with
perturbative theory of this
"field theory". Now let us return from these philosophical discussiones.
It should be remembered constituent quark is only virtual field here. In
real world, its kinetic degree of freedom is contained by its composited
states, e.g., the lowest order meson resonances and nucleon. Since scalar
octet and singlet of chiral SU(3) are not confirmed by experiment, we will
ignore them in this paper.

According to provious discussion on spin-1 meson resonances, the MG model
is easily to extended to include spin-1 meson resonances,
\begin{eqnarray}\label{2.16}
{\cal L}_{\chi}&=&i\bar{q}\gamma\cdot(\pa+\Gamma
   +\tilde{g}_{_A}\Delta\gamma_5)q
    +\bar{q}\gamma\cdot(V+A\gamma_5)q-m\bar{q}q 
 -\frac{1}{4B_0}\bar{q}(\xi\chi^{\dag}\xi+\xi^{\dag}\chi\xi^{\dag})q
      \nonumber \\
 &&-\frac{1}{4B_0}\kappa\bar{q}(\xi\chi^{\dag}\xi
     -\xi^{\dag}\chi\xi^{\dag})\gamma_5q
   +\frac{F^2}{16}<\nabla_\mu U\nabla^\mu U^{\dag}> 
   +\frac{1}{4}m_0^2<V_\mu V^{\mu}+A_\mu A^{\mu}>.
\end{eqnarray}
Although we have introduced current quark mass in lagrangian
(~\ref{2.16}), the pseudoscalar fields are still massless since
they are GoldStone bosons associated CSSB. The masses of
pseudoscalar mesons are generated via loop effects of constituent
quarks. In addition, $A_\mu-\pa^{\mu}\Phi$ mixing are also caused
by constituent quark loops. The symmetry requires this mixing to appear
according to form $<A_{\mu}\Delta^{\mu}>$. Thus this mixing can be
diagnolized via field shift
\begin{equation}\label{2.17}
 A_\mu\lraw A_\mu-ic\Delta_\mu.
\end{equation}
This field shift is nothing but to modify axial-vector current
coupling constant $\tilde{g}_{_A}$ in Eq.(~\ref{2.16}), i.e.,
$g_{_A}=\tilde{g}_{_A}-c$. Recalling axial-vector meson resonances
appear only as intermediate states, therefore, in fact, we can get rid of
axial-vector meson fields in lagrangian (~\ref{2.16}) and chiral
lagrangian (~\ref{2.16}) is rewirtten
\begin{eqnarray}\label{2.18}
{\cal L}_{\chi}&=&i\bar{q}\gamma\cdot(\pa+\Gamma+g_{_A}\Delta\gamma_5
    -iV)q-m\bar{q}q
  -\frac{1}{4B_0}\bar{q}(\xi\chi^{\dag}\xi+\xi^{\dag}\chi\xi^{\dag})q    
     \nonumber \\   
  &&-\frac{1}{4B_0}\kappa\bar{q}(\xi\chi^{\dag}\xi
     -\xi^{\dag}\chi\xi^{\dag})\gamma_5q
   +\frac{F^2}{16}<\nabla_\mu U\nabla^\mu U^{\dag}>
   +\frac{1}{4}m_0^2<V_\mu V^{\mu}>.
\end{eqnarray}
Here it should be rememberd that the experimental value $g_{_A}\simeq
0.75$ has included the effect of $A_\mu-\pa^{\mu}\Phi$ mixing.

\subsection{Effective lagrangian}

In this subsection, we like to derive the lowest order effective
lagrangian describing the coupling between vector meson resonances
and pseudoscalar mesons.

A review for chiral gauge theory is in \cite{Ball89}. The effective
lagrangian of mesons in ChQM can be obtained in Euclidian space by
means of integrating over degrees of freedom of fermions in
lagrangian(~\ref{2.18})
\begin{equation}
\label{2.19}
      \exp\{-\int d^4x{\cal L}_{eff}\}
      =\int {\cal D}\bar{q}{\cal D}q \exp\{-\int d^4x{\cal
       L}_\chi\}.
\end{equation}
Then we have
\begin{equation}
\label{2.20}
     {\cal L}_{eff}=-\ln{\rm det}{\cal D},
\end{equation}
with
\begin{equation}
\label{2.21}
{\cal D}=\gamma^\mu(\pa_\mu+\Gamma_\mu+g_A\Delta_\mu\gamma_5
  -iV_\mu)+m-\frac{1}{4B_0}(\xi\chi^{\dag}\xi+\xi^{\dag}\chi\xi^{\dag})
  -\frac{1}{4B_0}\kappa(\xi\chi^{\dag}\xi
    -\xi^{\dag}\chi\xi^{\dag})\gamma_5.
\end{equation}
The effective lagrangian is separated into two parts
\begin{eqnarray}
\label{2.22}
&&{\cal L}_{eff}={\cal L}_{eff}^{Re}+{\cal L}_{eff}^{Im}
       \nonumber   \\
&&{\cal L}_{eff}^{Re}=-\frac{1}{2}\ln{\rm det}({\cal D}{\cal D}^{\dag}),
 \hspace{0.8in}
 {\cal L}_{eff}^{Im}=-\frac{1}{2}\ln{\rm det}[({\cal D}^{\dag})^{-1}
     {\cal D}]
\end{eqnarray}
where
\begin{equation}
\label{2.23}
 {\cal D}^{\dag}=\gamma_5\hat{{\cal D}}\gamma_5,
\end{equation}
and $\hat{B}=\frac{1}{2}(1+\gamma_5)B_L+\frac{1}{2}(1-\gamma_5)B_R$
for arbitrarily operator
$B=\frac{1}{2}(1-\gamma_5)B_L+\frac{1}{2}(1+\gamma_5)B_R$. The
effective lagrangian ${\cal L}_{eff}^{Re}$ describes the physical
processes with normal parity and ${\cal L}_{eff}^{Im}$ the
processes with anomal parity. In the present paper we focus our attention
on $ {\cal L}_{eff}^{Re}$. The discussion of ${\cal L}_{eff}^{Im}$ can be
found in Refs.\cite{Ball89,Li95a}. In terms of Schwenger's proper time
method \cite{Schw54}, ${\cal L}_{eff}^{Re}$ is written as
\begin{equation}
\label{2.24}
  {\cal L}_{eff}^{Re}=-\frac{1}{2\delta(0)}\int d^4x\frac{d^4p}{(2\pi)^4}
   Tr\int_0^{\infty}\frac{d\tau}{\tau}(e^{-\tau{\cal D}^{\prime\dagger}
   {\cal D}^\prime}-e^{-\tau\Delta_0})\delta^4(x-y)|_{y\rightarrow x}
\end{equation}
with
\begin{eqnarray}
\label{2.25}
   &&{\cal D}^\prime={\cal D}-i\gamma \cdot p,  \hspace{0.8in}
   {\cal D}^{\prime\dag}={\cal D}^{\dag}+i\gamma\cdot p, \nonumber \\
   &&\Delta_0=p^2+M^2.
\end{eqnarray}
where M is an arbitrary parameter with dimension of mass. The
Seeley-DeWitt coefficients or heat kernel method have been used to evaluate
the expansion series of Eq.(~\ref{2.25}). In this paper we
will use dimensional regularization.
After completing the integration over $\tau$,
the lagrangian ${\cal L}_{eff}^{Re}$ reads
\begin{equation}
\label{2.26}
    {\cal L}_{eff}^{Re}=-\frac{\mu^\ep}{2\delta(0)}\int d^Dx
        \frac{d^Dp}{(2\pi)^D}\sum_{i=1}^{\infty}\frac{1}{n\Delta_0^n}
         Tr({\cal D}^{\prime\dagger}{\cal D}^\prime-\Delta_0)^n
         \delta^D(x-y)|_{y\rightarrow x},
\end{equation}
where trace is taken over the color, flavor and Lorentz space. This
effective lagrangian can be expanded in powers of derivatives,
\begin{equation}\label{2.27}
{\cal L}_{Re}={\cal L}_2+{\cal L}_4+....
\end{equation}
At order $p^2$, we will encounter logarithmic and quadratic divergences in
effective lagrangian generated by quark loops. The logarithmic divergence
can be canceled via renormalization of kinetic term of pseudoscalar
mesons. However, the quadratic divergence can not be renormalized. Thus we
need to define a constant $B_0$ to factorize the quadratic divergence(or
equivalently, to introduce a cut-off to truncate the divergence).
Explicitly, ${\cal L}_2$ reads
\begin{equation}\label{2.28}
{\cal L}_2=\frac{F_0^2}{16}<\na_\mu U\na^{\mu}U^{\dag}+\chi
  U^{\dag}+U\chi^{\dag}>+\frac{1}{4}m_0^2<V_\mu V^{\mu}>,
\end{equation}
where
\begin{eqnarray}\label{2.29}
\frac{F_0^2}{16}&=&\frac{F^2}{16}+\logdiv{}g_A^2m^2, \nonumber \\
\frac{F_0^2}{16}B_0&=&\quadiv{}m^3.
\end{eqnarray}
In chiral limit, it is known that $F_0$ just is decay constants of
$\pi$ mesons, $F_0=f_\pi=185$MeV. The lagrangian (~\ref{2.28}) yields
equation of motion of pseudoscalar mesons
\begin{equation}\label{2.30}
 \na_\mu(U\na^\mu U^{\dag})+\frac{1}{2}(\chi U^{\dag}-U\chi^{\dag})=0.
\end{equation}
Up to order $p^4$, all pseudoscalar meson fields satisfy this equation.

At order $p^4$ the effective lagrangian generated by quark loops reads
\begin{eqnarray}\label{2.31}
{\cal L}_{4}^{(q)}&=&-[\frac{g^2}{8}-\frac{\gamma}{12}]
   <L_{\mu\nu}L^{\mu\nu}+R_{\mu\nu}R^{\mu\nu}>
   -\frac{\gamma}{6}<L_{\mu\nu}R^{\mu\nu}> \nonumber \\
 &&-\frac{i\gamma}{3}g_A^2<\na_{\mu}U\na_{\nu}U^{\dag}
 \xi 
 R^{\mu\nu}\xi^{\dag}+\na_{\mu}U^{\dag}\na_{\nu}U\xi^{\dag}L^{\mu\nu}\xi>
  +\frac{\gamma}{12}g_A^4<\na_{\mu}U\na_{\nu}U^{\dag}
       \na^{\mu}U\na^{\nu}U^{\dag}>
     \nonumber \\
&&+\theta_1g_A^2<\na_\mu U\na^\mu U^{\dag}(\chi
     U^{\dag}+\chi^{\dag}U)>
  +\theta_2<\chi U^{\dag}\chi U^{\dag}+\chi^{\dag}U\chi^{\dag}U>
\end{eqnarray}
where
\begin{eqnarray}\label{2.32}
 \frac{3}{8}g^2&=&\logdiv{},\hspace{0.5in}
\gamma=\frac{N_c}{(4\pi)^2},\hspace{0.5in}
\theta_1=(\frac{3}{8}g^2-\gamma)\frac{m}{2B_0}, \nonumber \\
\theta_2&=&\frac{F_0^2}{128B_0m}(3-\kappa^2)+
\frac{3m}{64B_0}g^2(\frac{m}{B_0}-\kappa g_{_A}+\frac{g_{_A}^2}{2})
     -\frac{\gamma}{24}g_A^2, \\
L_{\mu\nu}&=&\frac{1}{2}(1+g_A)\xi F_{\mu\nu}^L\xi^{\dag}
        +\frac{1}{2}(1-g_A)\xi^{\dag}F_{\mu\nu}^R\xi+V_{\mu\nu}
        -i(1-g_A^2)[\Delta_\mu,\Delta_\nu]  
        -g_A([\Delta_\mu,V_\nu]+[V_\mu,\Delta_\nu]), \nonumber \\
R_{\mu\nu}&=&\frac{1}{2}(1+g_A)\xi^{\dag}F_{\mu\nu}^R\xi
        +\frac{1}{2}(1-g_A)\xi F_{\mu\nu}^L\xi^{\dag}+V_{\mu\nu}
        -i(1-g_A^2)[\Delta_\mu,\Delta_\nu]  
        +g_A([\Delta_\mu,V_\nu]+[V_\mu,\Delta_\nu]), \nonumber
\end{eqnarray}
with
\begin{eqnarray}\label{2.33}
F_{\mu\nu}^{R.L}&=&\partial_\mu(v_\nu\pm a_\nu)
      -\partial_\mu(v_\nu\pm a_\nu)-i[v_\mu\pm a_\mu,v_\nu\pm a_\nu].
         \nonumber \\
V_{\mu\nu}&=&d_{\mu}V_\nu-d_{\nu}V_\mu
   -i[V_\mu,V_\nu].
\end{eqnarray}
Here an universal coupling constant $g$ of the effective field theory
absorbs logarithmic divergences in Eq.~(\ref{2.31}).

From the kinetic terms of meson fields in lagrangians (~\ref{2.28}) and
(~\ref{2.31}) we can see that meson fields are not physical. The physical
meson fields can be defined via the following field rescaling in effective
lagrangian which make the kinetic terms of meson fields into the standard
form
\begin{equation}\label{2.34}
V_\mu\lraw\frac{1}{g}V_\mu,\hspace{1in}\Phi\lraw\frac{2}{f_\Phi}\Phi.
\end{equation}
where $\Phi=\pi,\;K,\;\eta$.

\subsection{Vector meson dominant and KSRF sum rules}

The direct coupling between photon and vector meson resonances is
also yielded by the effects of quark loops. Therefore, if vector
meson resonances are treated as bound states of constituent quarks,
vector meson dominant will be yielded naturally instead of input.
At isovector channel it reads from lagrangian (~\ref{2.31})
\begin{equation}\label{2.35}
  {\cal L}_{\rho\gamma}=-\frac{1}{4}eg(\pa^\mu{\cal A}^\nu-\pa^\nu{\cal A}^\mu)
  (\pa_\mu\rho_\nu^0-\pa_\nu\rho_\mu^0),
\end{equation}
where ${\cal A}_\mu$ is photon fields. Above equation is just the expression of
VMD proposed by Sakurai\cite{Sakurai69}. Similarly, at isoscalar channel they
read
\begin{eqnarray}\label{2.36}
 {\cal L}_{\omega\gamma}&=&-\frac{1}{12}eg(\pa^\mu{\cal A}^\nu
   -\pa^\nu{\cal A}^\mu)(\pa_\mu\omega_\nu-\pa_\nu\omega_\mu),\nonumber \\
 {\cal L}_{\phi\gamma}&=&\frac{1}{6}eg(\pa^\mu{\cal A}^\nu
   -\pa^\nu{\cal A}^\mu)(\pa_\mu\phi_\nu-\pa_\nu\phi_\mu).
\end{eqnarray}

It is well known that the KSRF(I) sum rule\cite{KSRF}
\begin{equation}\label{2.37}
 g_{\rho\gamma}(q^2)=\frac{1}{2}f_{\rho\pi\pi}(q^2)f_\pi^2
\end{equation}
is the result of current algebra and PCAC. So that it is expected
to be available at the leading order of momentum expansion. The
$g_{\rho\gamma}(q^2)$ is obtained from experssion (~\ref{2.35})
\begin{equation}\label{2.38}
g_{\rho\gamma}(q^2)=\frac{1}{2}gq^2
\end{equation}

In addition, if we set $m_u=m_d=0$, the lowest order
$\rho\rightarrow\pi\pi$ vertex reads
\begin{eqnarray}\label{2.39}
{\cal L}_{\rho\pi\pi}&=&f_{\rho\pi\pi}(q^2)\epsilon^{ijk}\rho_i^\mu
   \pi_j\pa_\mu\pi_k, \nonumber \\
f_{\rho\pi\pi}(q^2)&=&\frac{q^2}{gf_\pi^2}[g^2-(g^2
   -\frac{N_c}{3\pi^2})g_A^2].
\end{eqnarray}
From Eqs.~(\ref{2.38}) and (\ref{2.39}) we can find that
$g=\sqrt{\frac{N_c}{3}}\frac{1}{\pi}$ satisfy KSRF(I) sum rule
exactly. Therefore, $g\equiv\sqrt{\frac{N_c}{3}}\frac{1}{\pi}$
(especially, $g\equiv\pi^{-1}$ for $N_c=3$) is favorite choice for the
universal constant of the model. 

The interaction of vector meson resonances in the effective 
lagrangian~(\ref{2.31}) is similar to WCCWZ lagrangian given by
Brise\cite{Brise97}.
It is of an expected result since the symmetry realization of vector
mesons in our model is the same as one in WCCWZ approch. In addition,
those phenomenological requirements, such as VMD and univesal coupling are
also satisfied in this lowest order effective lagrangian. It implies that
ChCQM is legitimate approch on vector meson resonances. However, the above
$f_{\rho\pi\pi}(m_\rho^2)$ and $g_{\rho\gamma}(m_\rho^2)$ yield that the
widths of two on-shell decays are $\Gamma(\rho\rightarrow\pi\pi)=125$MeV
and $\Gamma(\rho\rightarrow e^+e^-)=4.35$KeV. Comparing with experiment
data, the error bars of those theoretical widths are about $15\%$ and
$35\%$ respectively. It can be naturally understood since the
contributions from high order terms of momentum expansion are droped here.
Thus we expect that these droped contributions could make the theoretical
prediction close to data.

\subsection{Low energy limit}

It is well known that, at very low energy, ChPT is a rigorous consequence
of the symmetry pattern of QCD and its spontaneous breaking. So that the
low energy limit of ChCQM must match with ChPT. The low energy limit of
this model can be obtained via integrating over vector meson resonances.
It means that, at very low energy, the dynamics of vector mesons are
replaced by pseudoscalar meson fields. Since there are no interaction of
vector mesons in ${\cal L}_2$, at very low energy, the equation of motion
$\delta{\cal L}/\delta V_\mu=0$ yields classics solution for vector mesons
are follow
\begin{equation}\label{2.40}
V_\mu=\frac{1}{m_{_V}^2}\times O(p^3){\rm terms},
\end{equation}
where $p$ is momentum of pseudoscalar at very low energy.
Therefore, in lagrangian (~\ref{2.31}), the terms involving vector
meson resonances are $O(p^6)$ at very low energy and do not
contribute to $O(p^4)$ low energy coupling constants,
$L_i(i=1,2,...,10)$. The low energy coupling constants yielded by
ChCQM(besides of $L_7$) can be directly obtained from
lagrangian~(\ref{2.31})
\begin{eqnarray}\label{2.41}
L_1&=&\frac{1}{2}L_2=\frac{\gamma}{24}, \hspace{0.8in}
L_3=-\frac{\gamma}{4}+\frac{\gamma}{12}g_A^4, \nonumber \\
L_4&=&L_6=0, \hspace{1in}L_5=\frac{\gamma m}{2B_0}g_A^2\\
L_8&=&\frac{F_0^2}{128B_0m}(3-\kappa^2)+       
\frac{3m}{64\pi^2B_0}(\frac{m}{B_0}-\kappa g_{_A}+\frac{g_{_A}^2}{2})
     -\frac{1}{128\pi^2}g_A^2, \nonumber \\
L_9&=&\frac{\gamma}{3}, \hspace{1.35in}
L_{10}=-\frac{\gamma}{3}+\frac{\gamma}{6}g_A^2. \nonumber
\end{eqnarray}

The constants $L_7$ has been known to get dominant contribution
from $\eta_0$\cite{GL85a} and this contribution is suppressed by
$1/N_c$. If we ignore the $\eta-\eta^{\prime}$ mixing, we have
\begin{equation}\label{2.42}
  L_7=-\frac{f_\pi^2}{128m_{\eta^{\prime}}^2}.
\end{equation}

Thus the five free parameters, $g$(it has been fitted by KSRF sum rule),
$g_{_A}$(it has been fitted by $n\rightarrow pe^-\bar{\nu}_e$
decay), $\kappa$, $m$ and $m_{\eta^{\pr}}$ determine
all ten low energy coupling constants of ChPT. It reflects the
dynamics constraints between those low energy coupling constants.
Here if we take $m_u+m_d\simeq 10$MeV, we can obtain
$B_0=\frac{m_\pi^2}{m_u+m_d}\simeq 2$GeV. Inputting experimental
values of $L_5$ and $L_8$, we obtain $m\simeq 480$MeV and $\kappa\simeq
0.2$. The numerical results for these low energy constants are in table 1.
We can find that all of them agree with experimental data well. Here the
constituent quark mass $m>m_\rho/2$, which is the same as out expectation.
In next section we will see that it is a necessary condition for yielding
a convergence expansion at $\rho$ mass scale.

\begin{table}[pb]
\centering
 \begin{tabular}{ccccccccccc}
&$L_1$&$L_2$&$L_3$&$L_4$&$L_5$&$L_6$&$L_7$&$L_8$&$L_9$&$L_{10}$
  \\ \hline
ChPT&$0.7\pm 0.3$&$1.3\pm 0.7$&$-4.4\pm 2.5$&$-0.3\pm 0.5$&$1.4\pm 0.5$& 
  $-0.2\pm 0.3$&$-0.4\pm 0.15$&$0.9\pm 0.3$&$6.9\pm 0.7$&$-5.2\pm 0.3$ \\
{ChCQM}&0.79&1.58&-4.25&0&$1.4^{a)}$&0&$(-0.4\pm 0.1)^{b)}$&
 $0.9^{a)}$&6.33&-4.55
   \end{tabular}
\begin{minipage}{5in}
\caption {\small $L_i$ in units of $10^{-3}$, ${\mu}=m_\rho$.
   a)input. b)contribution from gluon anomaly.}
\end{minipage}
\end{table}

\section{Diagram Analysis and Chiral Expansion at $\rho$ Mass Scale}
\setcounter{equation}{0}
\setcounter{figure}{0}

In previous section, we have derived the leading order effective
lagrangian of mesons via integrating out constituent quark fields in
original lagrangian. This path integral analysis is equivalent to
calculate the one-loop contribution of constituent quarks. The advantage
of path integral method is that we can derive an united effective
lagrangian of mesons, which is invariant under chiral
transformation for every orders of momentum expansion. However, it
is disadvantage of path integral method that it is hard to
calculate high order contribution of momentum expansion. This
shortage can be compensated through calculating one-loop graphs of
constituent quarks directly. In this section we will derive
effective vertices for $\gamma\rightarrow\pi\pi$, $\rho\rightarrow\pi\pi$,
4-pseudoscalar mesons and $\rho-\gamma$ coupling via diagram analysis.
All calculations will be performed at chiral limit.

We start with lagrangian(~\ref{2.18}). The effective action can be
obtained via integrating over degrees of freedom of fermions,
\begin{equation}\label{3.1}
e^{iS_{\rm eff}}\equiv\int{\cal D}\bar{q}{\cal D}qe^{i\int d^4x{\cal
   L}_\chi(x)}=<vac,out|in,vac>^{V,\Gamma,\Delta},
\end{equation}
where $<vac,out|in,vac>^{V,\Gamma,\Delta}$ is vacuum expectation value in 
presence of external source $V_\mu$, $\Gamma_\mu$ and $\Delta_\mu$. In
interaction picture, the above equation is rewritten as follow
\begin{eqnarray}\label{3.2}
e^{iS_{\rm eff}}&=&<0|{\cal T}_qe^{i\int d^4x{\cal L}^{\rm I}_\chi(x)}|0>
       \nonumber \\
 &=&\sum_{n=1}^\infty i\int d^4p_1\frac{d^4p_2}{(2\pi)^4}
  \cdots\frac{d^4p_n}{(2\pi)^4}\tilde{\Pi}_n(p_1,\cdots,p_n)
  \delta^4(p_1-p_2-\cdots-p_n) \nonumber \\
 &\equiv&i\Pi_1(0)+\sum_{n=2}^\infty i\int\frac{d^4p_1}{(2\pi)^4}
  \cdots\frac{d^4p_{n-1}}{(2\pi)^4}\Pi_n(p_1,\cdots,p_{n-1}),
\end{eqnarray}
where ${\cal T}_q$ is time-order product of constituent quark fields,
${\cal L}_{\chi}^{\rm I}$ is interaction part of lagrangian(~\ref{2.18}),
$\tilde{\Pi}_n(p_1,\cdots,p_n)$ is one-loop effects of constituent quarks
with $n$ external sources(hereafter we call it as $n$-point effective
vertex in momentum space), $p_1,p_2,\cdots,p_n$ are momentums of n
external sources respectively and
\begin{equation}\label{3.3}
\Pi_n(p_1,\cdots,p_{n-1})=\int d^4p_n\tilde{\Pi}_n(p_1,\cdots,p_n)
  \delta^4(p_1-p_2-\cdots-p_n).
\end{equation}
To get rid of all disconnected diagrams, we have
\begin{eqnarray}\label{3.4}
S_{\rm eff}&=&\int d^4x{\cal L}_{\rm eff}(x)=\Pi_1(0)
  +\sum_{n=2}^\infty\int \frac{d^4p_1}{(2\pi)^4}
  \cdots\frac{d^4p_{n-1}}{(2\pi)^4}\Pi_n(p_1,\cdots,p_{n-1})
         \nonumber \\
&\Rightarrow&{\cal L}_{\rm eff}(x)=\sum_{n=1}^\infty\int
  d^4p_1\frac{d^4p_2}{(2\pi)^4}\cdots\frac{d^4p_n}{(2\pi)^4}
  e^{i(p_1-p_2-\cdots-p_n)\cdot x}\tilde{\Pi}_n(p_1,\cdots,p_n).
\end{eqnarray}

\subsection{Two-point effective vertex}

There is no tapole diagram contribution of fermions, i.e., $\Pi_1(0)\equiv
0$. Thus we start calculating two-point effective vertex $\Pi_2(p)$
generated by fermion loop in figure(3.1). Due to parity conservation, here
both of two external sources are vector external sources
$(V_\mu+i\Gamma_\mu)$, or axial-vector external sources $g_A\Delta_\mu$.

\begin{figure}[hptb]
\label{vv}
   \centerline{\psfig{figure=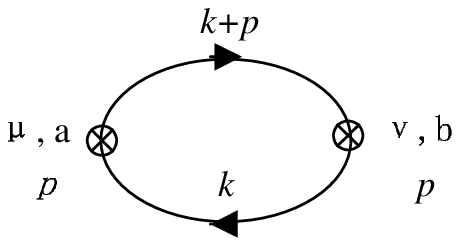,width=2.5in}}
 \centering
\begin{minipage}{5in}
   \caption{Two-point effective vertex generated by constituent
   quark one-loop, where $p$ is external momentum and $k$ is momentum of
   interal lines.}
\end{minipage}
\end{figure}

Employing the completeness relation of generators
$\lambda^a(a=1,2,...,N^2-1)$ of SU(N) group
\begin{eqnarray}\label{3.5}
&&<\lambda^aA\lambda^aB>=-\frac{2}{N}<AB>+2<A><B>,\nonumber \\
&&<\lambda^aA><\lambda^aB>=2<AB>-\frac{2}{N}<A><B>,
\end{eqnarray}
the two-point effective action is easily to obtain
\begin{eqnarray}\label{3.6}
S_2&=&\frac{f_\pi^2}{16}\int d^4x<\na_\mu U\na^\mu U^{\dag}>
   \nonumber \\
&&-\int\frac{d^4p}{(2\pi)^4}\frac{A(p^2)}{4}(\delta_{\mu\nu}p^2
  -p_{\mu}p_{\nu})<(V^{\mu}(p)+i\Gamma^{\mu}(p))
  (V^{\nu}(-p)+i\Gamma^{\nu}(-p))>,
\end{eqnarray}
where
\begin{equation}\label{3.7}
A(p^2)\equiv g^2-\frac{N_c}{\pi^2}\int_0^1dx\cdot x(1-x)\ln{(1-
   \frac{x(1-x)p^2}{m^2})}.
\end{equation}

From Eq.(~\ref{3.7}) we can see that unitarity of the effective
theory requires $4m^2>p^2$. This requirement also ensures that the
momentum expansion is convergent. Here we must point out that we
can not work in chiral limit simply if we want to study on-shell
$K^*(892)$ physics or on-shell $\phi(1020)$ physics. The reason is
that $4m^2>p^2\sim m_\rho^2$ can not ensure $4m^2>p^2\sim
m_{K^*}^2$ or $4m^2>p^2\sim m_\phi^2$. If we set $m_u=m_d=0$ but
$m_s\neq 0$, the unitarity and convergence of momentum expansion
require that $4m(m+m_s)>p^2$ for $p^2\simeq m_{K^*}^2$, and
$4(m+m_s)^2>p^2$ for $p^2\simeq m_{\phi}^2$. Those requirements are
satisfied indeed for $4m^2>m_\rho^2$ and usual value of strange quark
mass, i.e., $m_s\sim 150$MeV. Therefore, strange quark mass
can not be omitted when we study chiral expansion in $m_{_{K^*}}$ or
$m_\phi$ scale. In this paper, since we work in $\rho$-meson scale,
chiral limit is a good approximation.

In the following, we derive those effective vertices from Eq.(~\ref{3.6})
which relate to the purposes of this paper. The free field lagrangian of
$\rho$-meson reads
\begin{equation}\label{3.8}
{\cal L}_{\rm kin}^{(\rho)}=-\frac{1}{4}\rho_{\mu\nu}^i\rho^{i\mu\nu}
     +\frac{1}{2}m_\rho^2\rho_\mu^i\rho^{i\mu}.
\end{equation}
where $\rho_{\mu\nu}^i=(\pa_\mu\rho_\nu^i-\pa_\nu\rho_\mu^i)$. The above
lagrangian yields the classic equation of motion of $\rho$-meson in
momentum space as follow
\begin{equation}\label{3.9}
\frac{1}{2}(p^2\delta_{\mu\nu}-p_\mu
 p_\nu-m_\rho^2\delta_{\mu\nu})\rho^\mu=0
\end{equation}
Since in the present paper, all $\rho$-fields are treated at tree level,
they should obey the above equation of motion.

The two-point vertex $\rho$-meson reads
\begin{equation}\label{3.10}
{\cal L}_{\rho\rho}=-\frac{1}{4}[\frac{A(p^2)}{g^2}-1]
 \rho_{\mu\nu}^i\rho^{i\mu\nu},
\end{equation}
where $p^2$ defined by $p^2\rho_\nu=-\pa^2\rho_\nu$ is operator in
coordinate space. Using this equation of motion in Eq.(~\ref{3.9}), we
have
\begin{equation}\label{3.11}
m_\rho^2=\frac{m_0^2}{g^2}+\frac{N_c}{\pi^2g^2}m_\rho^2\int_0^1dx\cdot
   x(1-x)\ln{(1-\frac{x(1-x)m_\rho^2}{m^2})}.
\end{equation}
The experimental data $m_\rho=770$MeV yields $m_0=288$MeV.

The effective vertex describing $\rho-\gamma$ coupling reads
\begin{equation}\label{3.12}
{\cal L}_{\rho\gamma}=-\frac{1}{8}A(p^2)
  <\rho_{\mu\nu}(\xi\gamma^{\mu\nu}\xi^{\dag}
  +\xi^{\dag}\gamma^{\mu\nu}\xi)>,
\end{equation}
where
\begin{equation}\label{3.13}
\gamma_{\mu\nu}=e{\cal Q}(\pa_\mu{\cal A}_\nu-\pa_\nu{\cal A}_\mu),
\end{equation}
with ${\cal Q}={\rm diag}\{2/3,-1/3,-1/3\}$ is charge operator of
quark fields. Comparing effective vertices (~\ref{3.12}) with
leading order effective lagrangian (~\ref{2.31}), we can see that
the couplings in Eqs.(~\ref{3.12}) is momentum-dependent. 

\subsection{Contribution of triangle diagram}

There are two triangle diagrams(figure (3.2)) which also concern the
4-pseudoscalar meson, $\gamma\rightarrow\vphi\vphi$ and
$\rho\rightarrow\vphi\vphi$ vertices. The calculation on figure (3.2) is
well-known,
\begin{eqnarray}\label{3.14}
\Pi_3(p,q)=-\frac{1}{2}g_A^2B(p^2)p^\mu 
 <(V_\nu(p)+i\Gamma_\nu(p))[\Delta_\mu(p-q),\Delta_\nu(q)]>,
\end{eqnarray}
where
\begin{equation}\label{3.15}
B(p^2)=-g^2+\frac{N_c}{2\pi^2}\int_0^1dx\cdot x\int_0^1dy(1-xy)
   [1+\frac{m^2}{m^2-f(p^2)}+\ln{(1-\frac{f(p^2)}{m^2})}],
\end{equation}
where $g^2$ absorbes the logarithmic divergence from loop integral,
and $f(p^2)=x(1-x)(1-y)p^2$. 

\begin{figure}[hptb]
\label{trip}
   \centerline{\psfig{figure=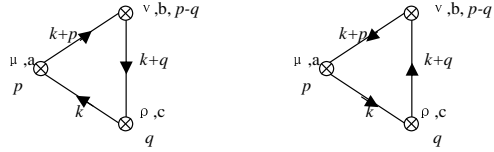,width=3in}}
  \centering
\begin{minipage}{5in}
   \caption{The triangle diagrams relate to $\gamma\rightarrow\vphi\vphi$
   and $\rho\rightarrow\vphi\vphi$, where $p$ is external momentum of
   vector current, $q$ is external momentum of axial-vector current and
   $k$ is momentum of interal lines.}
\end{minipage}
\end{figure}

Since high order diagrams(e.g., box diagram) do not contribute to
$\rho-\gamma$, $\gamma\rightarrow\vphi\vphi$ and
$\rho\rightarrow\vphi\vphi$ vertices, we do not calculate them here.
Then due to Eqs.(~\ref{3.6}) and (~\ref{3.14}), the effective lagrangian
describing vector-$\vphi\vphi$ vertices is follow
\begin{eqnarray}\label{3.16}
{\cal L}_{V\pi\pi}&=&-\frac{i}{16}gf_\pi^2b(p^2)<(\rho_{\mu\nu}
   (\xi\na^{\mu}U^{\dag}\na^{\nu}U\xi^{\dag}
   +\xi^{\dag}\na^{\mu}U\na^{\nu}U^{\dag}\xi) \nonumber \\
   &&+\gamma_{\mu\nu}
   (\na^{\mu}U^{\dag}\na^{\nu}U+\na^{\mu}U\na^{\nu}U^{\dag})>.
\end{eqnarray}
where
\begin{equation}\label{3.17}
b(p^2)=\frac{1}{gf_\pi^2}[A(p^2)+g_A^2B(p^2)].
\end{equation}

Similarly, since
$$\Gamma_{\mu\nu}=-\frac{i}{2}(\xi\gamma_{\mu\nu}\xi^{\dag}
 +\xi^{\dag}\gamma_{\mu\nu}\xi)-[\Delta_\mu,\Delta_\nu],$$
the quark-loop effects from figure(3.1) and (3.2) also contribute to four
pseudoscalar meson vertex. The result is
\begin{eqnarray}\label{3.18}
{\cal L}'_{4P}&=&\frac{1}{16}f_\pi^2C(p^2)<\Omega_{\mu\nu}
  (\xi\na^{\mu}U^{\dag}\na^{\nu}U\xi^{\dag}
   +\xi^{\dag}\na^{\mu}U\na^{\nu}U^{\dag}\xi)> \nonumber \\
&=&\frac{1}{64}
  f_\pi^2C(p^2)<\na_\mu U\na_\nu U^{\dag}\na^\mu U\na^\nu U^{\dag}
  -\na_\mu U\na^\mu U^{\dag}\na_\nu U\na^\nu U^{\dag}>,
\end{eqnarray}
where
\begin{eqnarray}\label{3.19}
C(p^2)&=&=\frac{1}{2f_\pi^2}[A(p^2)+2g_{_A}^2B(p^2)] \nonumber \\
\Omega_\mu&=&\frac{1}{2}(\xi\pa_\mu\xi^{\dag}+\xi^{\dag}\pa_\mu\xi),
 \nonumber \\
\Omega_{\mu\nu}&=&\pa_\mu\Omega_\nu-\pa_\nu\Omega_\mu
 =-[\Delta_\mu,\Delta_\nu],
\end{eqnarray}
and $p$ is momentum operator of $\Omega$. In fact, the box diagram also
contributes to four-pseudoscalar meson vertex. However, in
Eq.(~\ref{2.41}) the $L_3$ tell us that this contribution is suppressed by
$g_{_A}^4/3\simeq 0.1$ at least. Thus here we omit the box diagram
contribution. Then Eq.(~\ref{3.18}) together with Eq.(~\ref{3.6}) include
all four-pseudoscalar vertices.

\subsection{What can break KSRF(I) sum rule}

Now let us calculate $\rho\rightarrow e^+e^-$ decay and
$\rho\rightarrow \pi\pi$ decay in which $\rho$-meson is
on-shell($p^2=m_\rho^2$). Taking $m=480$MeV which is fitted by
coupling constants of ChPT in sect. 2.4, we obtain $A(m_\rho^2)=0.139$
and $B(m_\rho^2)=-0.026$. Then we have $\Gamma(\rho\rightarrow
\pi\pi)=182$MeV and $\Gamma(\rho\rightarrow e^+e^-)=7.76$KeV. Recalling
the leading order results in sect. 2.3, 125MeV and 4.32KeV respectively,
we can see that the high order terms of the chiral expansion yield very
important contribution at $m_\rho$-scale. It is not surprising since we
have pointed out that the chiral expansion is slowly convergence in this
energy region. However, those theoretical values are much larger than
expertimental data 150MeV and 6.77KeV respectively. Thus how can we
understand these results? In addition, since $g_A^2B(m_\rho^2)$ is much
smaller than $A(m_\rho^2)$, the KSRF(I) sum rule is still kept well
here. It is well-known that theoretcial predictions of
$\Gamma(\rho\rightarrow\pi\pi)$ and $\Gamma(\rho\rightarrow e^+e^-)$ can
not match with data simultaneously if KSRF(I) sum rule is satisfied. Thus
what can break KSRF(I) sum rule? In next two sections we will show that,
contribution from meson loops also plays important role at this scale. It
provides a nature mechanism to break KSRF(I) sum rule, and makes
theoretical predicitions(on-shell decay of vertor mesons, form factor of
$\pi$, etc.) agree with experimental data very well.

\section{One-loop Graphs of Mesons}
\setcounter{equation}{0}
\setcounter{figure}{0}

A natural agruement is that the contribution from meson loops is
suppressed
by $N_c^{-1}$ expansion\cite{tH74}. However, for $N_c=3$ in real world,
this suppression is not large enough so that we can not omit the
contribution from meson loop. Moreover, the unitarity implies that the
imaginary part of ${\cal T}$-matrix is large at vector meson mass
scale, but the imaginary part of ${\cal T}$-matrix is generated by
meson loops only in this formalism. Thus at energy scale of vector meson
masses, the meson loop effects must be evaluated. Due to $N_c^{-1}$
expansion, the dominant contribution of meson loops is from one-loop
graphs. Furthermore, since there is large gap between vector meson mass
and pseudoscalar meson mass, the one-loop graphs of pseudoscalar mesons
yield the most important contribution here. This point is also shown from
that only the one-loop graphs of pseudoscalar mesons can yield the
imaginary part of ${\cal T}$-matrix in this energy region. Therefore, in
this section we will calculate one-loop effects of pseudoscalar mesons,
which correct $\rho-\gamma$, $\gamma\rightarrow\pi\pi$ and 
$\rho\rightarrow\pi\pi$ vertices. 

\begin{figure}[hptb]
\centerline{\psfig{figure=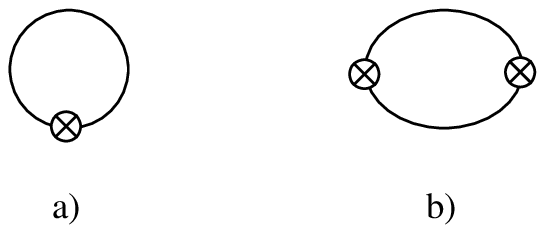,width=3in}}
 \centering
\begin{minipage}{5in}
   \caption{Two kinds of one-loop graphs generated by pseudoscalar mesons,
   which relate to calculation in this section. a) Tadpole diagram. b)
   Two-point corrector.}
\end{minipage}
\end{figure}

There are several remarks relating to our calculation, 1) Since
$m_\pi^2\ll m_{_K}^2<m_\rho^2$, we treat pion as massless particle but
$m_{_K}\neq 0$ in interal line. Moreover, since difference between
$m_{\eta_8}$ and $m_{_K}$ is small, we set $m_{\eta_8}=m_{_K}$ in this
section. 2) Since we focus our attention on $m_\rho$-energy scale, we set
current quark masses are zero in the following. 3) There are only
two diagrams relating to our calcuation on all potential irreducible
one-loop graphs. They are tadpole diagram(figure 4.1-a) and diagram
including one-loop with two external source(see figure(4.1-b), hereafter
we call contribution generate by this kinds of diagrams as two-point
corrector).

\subsection{Four pseudoscalar meson vertex}

The four pseudoscalar meson vertex relates to our following calculation
which can be obtained from Eqs.(~\ref{3.6}) and (~\ref{3.18}). Recalling
$m_\pi^2=0$ but $m_{\eta_8}^2=m_{_K}^2\neq 0$, we can see that only
$K$ and $\eta_8$ mesons can yields non-zero contribution of tadpole
diagram, since in dimensional regularization, $\int
d^{D}k(k^2+i\ep)^{-1}\equiv 0$. Obviously, the tadpole diagram contributes
a factor which is proportional to $m_{_K}^2$ and momentum-independent.
Thus tadpole-loop correction generated by ${\cal L}_2$ is nothing other
than renormalization of $f_\pi$. Here we calculate
the tadpole-loop correction generated by ${\cal L}'_{4P}$. The calculation
on two-point corrector of four pseudoscalar meson vertex will be included
in the following section so that we need not calculate it here.

It is convenient to calculate meson loops in terms of background field
method. To expand pseudoscalar meson fields around their classic solution
\begin{equation}\label{4.1}
U(x)=\bar{\xi}(x)e^{i\vphi}\bar{\xi}(x),\hspace{1in}
\bar{U}(x)=\bar{\xi}^2(\Phi),
\end{equation}
where background $\bar{U}(x)$ is solution of classic motion of
pseudoscalar mesons, $\delta{\cal L}/\delta U(x)=0$, $\vphi(x)$ is quantum
fluctuation fields around this classic solution. Inserting
Eq.(~\ref{4.1}) into the effective lagrangian in sect. 3 and retain
terms to quadratic form of quantum fields, the one-loop effects of
pseudoscalar meson can be obtained via integrating over ths quantum
fields. 

Then tadpole-loop contribution to 4-pseudoscalar meson vertex 
can be obtained via the following intregral
\begin{equation}\label{4.2}
\Pi_{\rm tad}^{(4P)}=i\int\frac{d^4k}{(2\pi)^4}\frac{i}
 {k^2-m_{_K}^2+i\ep}\frac{C(p^2)}{4}{\cal H}^{aa},
\end{equation}
where
\begin{equation}\label{4.3}
{\cal H}^{aa}[SU(3)/SU(2)]=\frac{11}{6}<\Omega_{\mu\nu}
  (\xi\na^{\mu}U^{\dag}\na^{\nu}U\xi^{\dag}
   +\xi^{\dag}\na^{\mu}U\na^{\nu}U^{\dag}\xi)>.
\end{equation}
Substituenting Eq.(~\ref{4.3}) into Eq.(~\ref{4.2}), we obtain the
tadpole-loop contribution to 4-pseudoscalar vertex as follow
\begin{equation}\label{4.4}
\Pi_{\rm tad}^{(4P)}=-\frac{11}{24}\frac{\lambda}{(4\pi)^2}
  m_{_K}^2C(p^2)<\Omega_{\mu\nu}(\xi\na^{\mu}U^{\dag}
  \na^{\nu}U\xi^{\dag}
   +\xi^{\dag}\na^{\mu}U\na^{\nu}U^{\dag}\xi)>.
\end{equation}
where we define a parameter $\lambda$ to absorbe
quadratic divergence from loop integral
\begin{equation}\label{4.5}
\lambda=(\frac{4\pi\mu^2}{m_{_K}^2})^{\ep/2}\Gamma(1-\frac{D}{2}).
\end{equation}

\begin{figure}[hptb]
\centerline{\psfig{figure=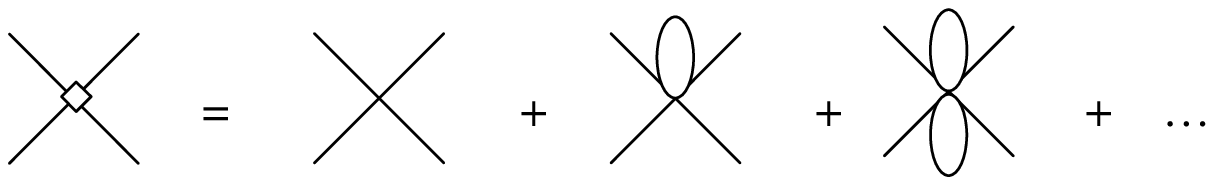,width=5.5in}}
  \centering
\begin{minipage}{5in}
   \caption{Four-pseudoscalar meson vertex with tadpole diagram
   correction. Here $``\diamond''$ denotes the vertex with correction of
   all tadpole diagrams.}
\end{minipage}
\end{figure}

For calculating all potential tadpole diagrams contribution, we need to
sum over all diagrams in figure(4.2). Comparing Eq.(~\ref{4.4}) with
Eq.(~\ref{3.18}), we can see every tadpole-loop in figure(4.2) contributes
a factor
\begin{equation}\label{4.6}
-\frac{11}{3}\zeta,\hspace{1in}
(\zeta=\frac{2\lambda}{(4\pi)^2}\frac{m_{_K}^2}{f_\pi^2}).
\end{equation}
Then to sum over all diagrams in figure(4.2), we obtain
\begin{equation}\label{4.7}
{\cal L}'_{4P}=\frac{1}{16}\frac{f_\pi^2C(p^2)}{1+11\zeta/3}
<\Omega_{\mu\nu}(\xi\na^{\mu}U^{\dag}\na^{\nu}U\xi^{\dag}
   +\xi^{\dag}\na^{\mu}U\na^{\nu}U^{\dag}\xi)>.
\end{equation}

For the sake of convenience of calculation on two-point corrector in the
following subsections, we like to divid quantum pseudoscalar fields from
lagrangian ${\cal L}_2$ and ${\cal L}'_{4P}$ in terms of background field
method. Those quantum pseudoscalar fields contract to internal lines in
figure(4.1-b).

Inserting Eq.(~\ref{4.1}) in to lagrangian ${\cal L}_2$ and retaining
terms to quadratic form of quantum fields we obtain
\begin{equation}\label{4.8}
{\cal L}_2=\bar{{\cal L}}_2+\frac{f_\pi^2}{16}<d_\mu\vphi d^{\mu}\vphi
   -[\Delta_\mu,\vphi][\Delta^\mu,\vphi]>.
\end{equation}
Then interaction vertices including two quantum fields is
\begin{equation}\label{4.9}
\delta{\cal L}_2=\frac{1}{4}<2\Gamma_\mu[\vphi,\pa^\mu\vphi]
  +[\Gamma_\mu,\vphi][\Gamma^\mu,\vphi]-[\Delta_\mu,\vphi]
   [\Delta^\mu,\vphi]>,
\end{equation}
where quantum field has been normalized, and only background fields
are included in $\Gamma_\mu$ and $\Delta_\mu$,
\begin{equation}\label{4.10}
\Gamma_\mu=-i\gamma_\mu+\Omega_\mu=-i\gamma_\mu
  +\frac{1}{8}[\Phi,\pa_\mu\Phi]+...,
  \hspace{1in}\gamma_\mu=e{\cal Q}A_\mu,
\end{equation}
where $\Phi$ is background fields. In lagrangian(~\ref{4.9}), only first
term relates to our following calculation.

Inserting Eq.(~\ref{4.1}) into Eq.(~\ref{4.7}) and retaining
terms to quadratic form of quantum fields we have
\begin{equation}\label{4.11}
\delta{\cal L}'_{4P}(p^2)=\frac{1}{2}\frac{C(p^2)}{1+11\zeta/3}
  <\Omega_{\mu\nu}[\pa^{\mu}\vphi,\pa^\nu\vphi]>,
\end{equation}
where $p$ is momentum of $\Omega_\mu$ and independent of loop integral.
This is only term which survives when coupled to conserved current or
on-shell vector mesons. Eq.(~\ref{4.11}) together with Eq.(~\ref{4.9})
lead to all 4-pseudoscalar meson vertex which relates to our the
following calculation,
\begin{equation}\label{4.12}
\delta{\cal L}_{4P}=\frac{1}{2}[\delta_{\mu\nu}+\frac{C(p^2)}{1+11\zeta/3}
 (p^2\delta_{\mu\nu}-p_\mu p_\nu)]<\Omega^{\mu}[\vphi,\pa^\nu\vphi]>.
\end{equation}

\subsection{Correction to $\gamma\rightarrow\pi\pi$ vertex}

\subsubsection{Tadpole diagram}

The effective lagrangian can generate non-trivial tadpole-loop
contribution, which corrects to $\gamma\rightarrow\vphi\vphi$ vertex(see
figure(4.3)). Obviously, the correction of tadpole diagram is proportional
to $m_{_K}^2$ and momentum-independent. Thus tadpole-loop correction
generated by ${\cal L}_2$ is nothing other than
renormalization of $f_\pi$. Here we calculate the tadpole-loop correction
generated by ${\cal L}'$. In terms of background method,
we can insert Eq.(~\ref{4.1}) into lagrangian (~\ref{3.16}) and retain
terms to quadratic form of quantum fields. Then we obtain
\begin{eqnarray}\label{4.13}
\delta{\cal L}_{\rm tad}&=&\frac{i}{2}b(p^2)
 <(\xi\gamma^{\mu\nu}\xi^{\dag}+\xi^{\dag}\gamma^{\mu\nu}\xi)
  (\Delta_\mu\vphi\Delta_\nu\vphi+\vphi\Delta_\mu\Delta_\nu\vphi
  +\vphi\Delta_\mu\vphi\Delta_\nu \nonumber \\ &&\hspace{1in}
  -\frac{3}{2}\Delta_\mu\Delta_\nu\{\vphi,\vphi\})>,
\end{eqnarray}
where $\vphi$ has been normailzed.

Due to completeness relation of generators of SU(N) group (~\ref{3.5}), we
have
\begin{eqnarray}\label{4.14}
\Pi_{\rm tad}^{\gamma\vphi\vphi}=\frac{3i}{4}\frac{\lambda}{(8\pi)^2}
   m_{_K}^2gb(p^2)<\gamma_{\mu\nu}(\na^{\mu}U^{\dag}\na^{\nu}U
      +\na^{\mu}U\na^{\nu}U^{\dag})>.
\end{eqnarray}

\begin{figure}[hptb]
\centerline{\psfig{figure=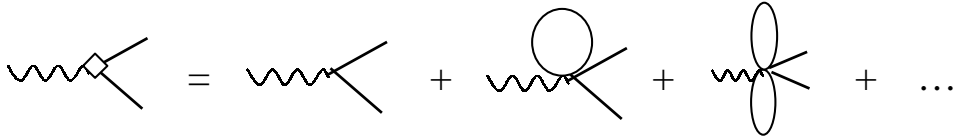,width=5.5in}}
 \centering
\begin{minipage}{5in}
   \caption{$\gamma\rightarrow\vphi\vphi$ vertex with correction of all
   tadpole diagrams.}
\end{minipage}
\end{figure}

Comparing Eq.(~\ref{4.14}) with Eq.(~\ref{3.16}), we can see that every
tadpole-loop in figure(4.3) yields a factor $-3\zeta$. To sum over all
diagrams in figure(4.3), we obtain the $\gamma\rightarrow\vphi\vphi$
vertex with tadpole diagram correction as follow
\begin{equation}\label{4.15}
{\cal L}^{(1)}_{\gamma\vphi\vphi}=-\frac{i}{2}
  <\gamma_\mu[\vphi,\pa^\mu\vphi]>-\frac{i}{4}\frac{gb(p^2)}{1+3\zeta}
 <\gamma_{\mu\nu}[\pa^\mu\vphi,\pa^\nu\vphi]>,
\end{equation}
where the quantum field $\vphi$ has been normalized.

\subsubsection{Two-point corretor}

\begin{description}
\item[a.]{$\gamma\rightarrow\vphi\vphi$ vertex generated by
   ${\cal L}_2$}
\end{description}

In this case, the tree level $\gamma\rightarrow\vphi\vphi(\vphi=\pi,K)$
vertex reads from the first term of Eq.(~\ref{4.15}), and 4-$\vphi$
vertices is given in Eq.(~\ref{4.12}). The calculation is straightforward
\begin{eqnarray}\label{4.16}
G_1(p)&=&\frac{i}{4}<\gamma_\mu[\lambda^a,\lambda^b]><\Omega^{\alpha}
  [\lambda^a,\lambda^b]>[\delta_{\alpha\beta}+\frac{C(p^2)}{1+11\zeta/3}
  (p^2\delta_{\alpha_\beta}-p_\alpha p_\beta)] \nonumber \\&&\times
  \int\frac{d^4k}{(2\pi)^4}\frac{i}{k^2-m_{\vphi}^2+i\ep}
  \frac{i}{(k+p)^2-m_{\vphi}^2+i\ep}k^\mu(2k+p)^\beta,
\end{eqnarray}
where $p$ is momentum of photon. Employing completeness relation of
generators of SU(N) group (~\ref{3.5}), we obtain
\begin{equation}\label{4.17}
<A[\lambda^a,\lambda^b]><B[\lambda^a,\lambda^b]>
=-8N<AB>+8<A><B>.
\end{equation}
Then recalling massless pion and U(1)$_{\rm e.m.}$ guage invariant, we
obtain SU(2) correction of Eq.(~\ref{4.16}) as follow
\begin{eqnarray}\label{4.18}
G_1[SU(2)]&=&\frac{i}{2}
  (p^2\delta_{\mu\nu}-p_\mu p_\nu)
  <\gamma^\mu[\Phi,\pa^\nu\Phi]>\{[\frac{\lambda}{96\pi^2}
   +\frac{1}{16\pi^2}\int_0^1dx\cdot x(1-x)\ln{\frac{x(1-x)p^2}
  {m_{_K}^2}} \nonumber \\
  &&+\frac{i}{96\pi^2}Arg(-1)\theta(p^2-4m_\pi^2)]
  -\frac{C(p^2)}{1+11\zeta/3}\frac{p^2}{(4\pi)^2}[\frac{\lambda}{6}
        \nonumber \\
  &&+\int_0^1dx\cdot x(1-x)\ln{\frac{x(1-x)p^2}{m_{_K}^2}}
    +\frac{i}{6}Arg(-1)\theta(p^2-4m_\pi^2)]\},
\end{eqnarray}
where
\begin{eqnarray}\label{4.19}
Arg(-1)&=&(1+2k)\pi,\hspace{1in}k=0,\pm 1,\pm 2,..., \nonumber \\
\theta(x-y)&=&\left\{
    {1;\hspace{0.8in} x>y \atop 0.\hspace{0.8in} x\leq y} \right.
\end{eqnarray}
We can see that one-loop of pion contributes to a large imaginary part of
${\cal T}$-matrix.

Since in this paper we pay our attention on energy scale
$p^2<4m^2\simeq 4m_{_K}^2$, there is no imaginary part yielded by
$K$-loop. Then we can obtain SU(3)/SU(2) correction of
Eq.(~\ref{4.16}) as follow
\begin{eqnarray}\label{4.20}
&&G_1[SU(3)/SU(2)]\hfill \nonumber \\
&=&-\frac{i}{4}(p^2\delta_{\mu\nu}-p_\mu p_\nu)
  <\gamma^\mu[\Phi,\pa^\nu\Phi]>
\{[\frac{\lambda}{96\pi^2}-\frac{1}{16\pi^2}\int_0^1\cdot x(1-x)
  \ln{(1-\frac{x(1-x)p^2}{m_{_K}^2})}] \nonumber \\
&&-\frac{C(p^2)}{1+11\zeta/3}\frac{1}{(4\pi)^2}[\lambda
  (\frac{p^2}{6}-m_{_K}^2)-\int_0^1dx\cdot[m_{_K}^2- x(1-x)p^2]
  \ln{(1-\frac{x(1-x)p^2}{m_{_K}^2})}]\}.
\end{eqnarray}

Defining
\begin{eqnarray}\label{4.21}
D(p^2)&=&\frac{1}{16\pi^2f_\pi^2}\{\lambda+\int_0^1dx\cdot
  x(1-x)\ln{[(1-\frac{x(1-x)p^2}{m_{_K}^2})
  (\frac{x(1-x)p^2}{m_{_K}^2})^2]}\nonumber \\&&\hspace{0.5in}
    +\frac{2}{3}Arg(-1)\theta(p^2-4m_\pi^2), \nonumber \\
\Sigma_0(p^2)&=&\frac{1}{8\pi^2f_\pi^2}\{\lambda(\frac{p^2}{2}
  -m_{_K}^2)-\int_0^1dx\cdot[m_{_K}^2-x(1-x)p^2]
   \ln{(1-\frac{x(1-x)p^2}{m_{_K}^2})}\nonumber \\
  &&+2p^2\int_0^1dx\cdot x(1-x)\ln{\frac{x(1-x)p^2}{m_{_K}^2}}
     +p^2\frac{i\pi}{3}Arg(-1)\theta(p^2-4m_\pi^2)\},
\end{eqnarray}
we obtain that two-point corrector of $\gamma\rightarrow\vphi\vphi$ vertex
which generated by ${\cal L}_2$ as follow
\begin{eqnarray}\label{4.22}
G_1^{\gamma\vphi\vphi}&=&\frac{i}{8}f_\pi^2
  [D(p^2)+\frac{C(p^2)\Sigma_0(p^2)}{1+11\zeta/3}]
  (p^2\delta_{\mu\nu}-p_\mu p_\nu)
  <\gamma^\mu[\Phi,\pa^\nu\Phi]> \nonumber \\
  &\rightarrow&\frac{i}{8}f_\pi^2
  [D(p^2)+\frac{C(p^2)\Sigma_0(p^2)}{1+11\zeta/3}]
 <\gamma_{\mu\nu}(\na^{\mu}U^{\dag}\na^{\nu}U
  +\na^{\mu}U\na^{\nu}U^{\dag})>.
\end{eqnarray}
This effective vertex is $O(p^4)$ at least.

\begin{description}
\item[b.]{$\gamma\rightarrow\vphi\vphi$ vertex generated by high
order lagrangian}
\end{description}

Since Eq.(~\ref{4.22}) and the second term of Eq.(~\ref{4.15}) are same
level in momentum expasion, we can calculate two-point corrector
generated by them simultaneously. Combining Eqs.(~\ref{4.22}) and
the second term of (~\ref{4.15}), we have
\begin{equation}\label{4.23}
{\cal L}'=-\frac{i}{2}b_\gamma(p^2)<\gamma_{\mu\nu}
    [\pa^\mu\vphi,\pa^\nu\vphi]>,
\end{equation}
where
\begin{equation}\label{4.24}
b_\gamma(p^2)=\frac{gb(p^2)}{2(1+3\zeta)}-D(p^2)
  -\frac{C(p^2)\Sigma_0(p^2)}{1+11\zeta/3}.
\end{equation}

Then we obtain two-point corrector generated by ${\cal L}'$ as follow
\begin{eqnarray}\label{4.25}
G_2^{\gamma\vphi\vphi}(p)&=&-\frac{N}{2}b_\gamma(p^2)
  <\gamma_{\mu\nu}[\Phi,\pa^\alpha\Phi]>
    [\delta_{\alpha\beta}+\frac{C(p^2)}{1+11\zeta/3}
  (p^2\delta_{\alpha_\beta}-p_\alpha p_\beta)]\nonumber \\ &&\times
    \int_0^1dx\int\frac{d^4k}{(2\pi)^4}\frac{p^\mu k^\nu k^\beta}
    {[k^2-m_{\vphi}^2+x(1-x)p^2+i\ep]^2}.
\end{eqnarray}
In case of SU(2), the above equation is rewritten as follow
\begin{eqnarray}\label{4.26}
G_2[SU(2)]&=&\frac{i}{2}b_\gamma(p^2)[1+\frac{p^2C(p^2)}{1+11\zeta/3}]
     <\gamma_{\mu\nu}[\pa^\mu\Phi,\pa^\nu\Phi]>\frac{p^2}{(4\pi)^2}
           \nonumber \\ &&\times
     \{\frac{\lambda}{6}+\int_0^1dx\cdot x(1-x)
     \ln{\frac{x(1-x)p^2}{m_{_K}^2}}
   +\frac{i}{6}Arg(-1)\theta(p^2-4m_\pi^2)\}.
\end{eqnarray}
For massive $K$-meson, two-point corrector in SU(3)/SU(2) sector reads
\begin{eqnarray}\label{4.27}
G_2[SU(3)/SU(2)]&=&\frac{i}{4}b_\gamma(p^2)
      [1+\frac{p^2C(p^2)}{1+11\zeta/3}]
      <\gamma_{\mu\nu}[\pa^\mu\Phi,\pa^\nu\Phi]>\frac{1}{(4\pi)^2}
      \{\lambda(\frac{p^2}{6}-m_{_K}^2) \nonumber \\
      &&-\int_0^1dx\cdot[m_{_K}^2
      -x(1-x)p^2]\ln{(1-\frac{x(1-x)p^2}{m_{_K}^2})}\}.
\end{eqnarray}
Eqs.(~\ref{4.27}) together with Eq.(~\ref{4.26}) given one-loop correction
to $\gamma\rightarrow\Phi\Phi$ vertex. Defining
\begin{equation}\label{4.28}
\Sigma(p^2)=(1+\frac{p^2C(p^2)}{1+11\zeta/3})\Sigma_0(p^2),
\end{equation}
we obtain
\begin{equation}\label{4.29}
G_2^{\gamma\vphi\vphi}=\frac{i}{8}b_\gamma(p^2)f_\pi^2\Sigma(p^2)
   <\gamma_{\mu\nu}[\pa^\mu\Phi,\pa^\nu\Phi]>.
\end{equation}

\begin{figure}[hptb]
\label{chain}
\centerline{\psfig{figure=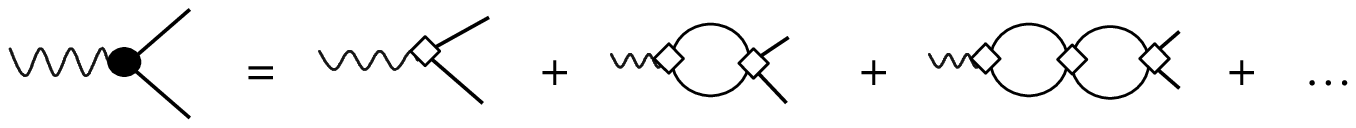,width=6in}}
\centering
\begin{minipage}{5in}
   \caption{The chain approximation. To sum over all of those diagrams we
   can obtain complete $\gamma\rightarrow\pi\pi$ vertex up to one-loop
   level. Here $``\diamond''$ denotes every vertices have included
   tadpole-loop correction.}
\end{minipage}
\end{figure}

To sum over all diagrams in chain approximation(figure(4.4)), we
can obtain the complete $\gamma\rightarrow\pi\pi$ vertex. Comparing
Eq.(~\ref{4.29}) with Eq.(~\ref{4.23}), we can see that every one-loop in
figure(4.4) yields a factor ($-\Sigma(p^2)$). Thus to sum over all
diagrams in figure(4.4) and together with the lowest order term, we can
obtain the complete $\gamma\rightarrow\pi\pi$ vertex as follow
\begin{equation}\label{4.30}
{\cal L}_{\gamma\pi\pi}^{c}=-\frac{i}{2}<\gamma_\mu[\pi,\pa^\mu\pi]>
  -\frac{i}{2}\frac{b_{\gamma}(p^2)}{1+\Sigma(p^2)}
  <\gamma_{\mu\nu}[\pa^\mu\pi,\pa^\nu\pi]>,
\end{equation}
where $\pi$ field has been normalized. The eq.~(\ref{4.30}) is important
for studies on $\omega$ physics and pion form factor in elsewhere.

\subsection{Correction to $\rho\rightarrow\pi\pi$ vertex}

In leading order of $N_c^{-1}$ expansion, the $\rho\rightarrow\vphi\vphi$
vertex read from(with physical $\rho$-field)
\begin{equation}\label{4.31}
{\cal L}_{\rho\vphi\vphi}=-\frac{i}{16}b(p^2)f_\pi^2
   <\rho_{\mu\nu}(\xi\na^\mu U^{\dag}\na^\nu U\xi^{\dag}
         +\xi^{\dag}\na^\mu U\na^\nu U^{\dag}\xi)>.
\end{equation}
Thus the calculation in this subsection is similar to one in section
4.2.2.

\begin{description}
\item[a.] {\bf Tadpole diagram}
\end{description}
In this case, $\xi$ and $\xi^{\dag}$ in lagrangian (~\ref{4.31}) include
quantum fields only, i.e., $\xi={\rm exp}\{i\vphi/2\}$. Then
inserting Eq.(~\ref{4.1}) and the above $\xi,\;\xi^{\dag}$ into lagrangian
(~\ref{4.31}), and retaining terms to quadratic form of quantum fields, we
obtain
\begin{eqnarray}\label{4.32}
\delta{\cal L}_{\rm tad}=
=\frac{N}{4}ib(p^2)\vphi^a\vphi^a<\rho_{\mu\nu}(\xi\na^\mu
    U^{\dag}\na^\nu U\xi^{\dag}+\xi^{\dag}\na^\mu U\na^\nu U^{\dag}\xi)>,
\end{eqnarray}
where quantum fields $\vphi^a$ have been normalized. From
Eq.(~\ref{4.32}), it is easily to obtain tadpole-loop contribution which
is yielded by pseudoscalar mesons of SU(3)/SU(2) sector,
\begin{eqnarray}\label{4.33}
\Pi_{\rm tad}^{\rho\vphi\vphi}=
  \frac{i}{4}\frac{\lambda}{(4\pi)^2}m_{_K}^2b(p^2)
   <\rho_{\mu\nu}(\xi\na^\mu U^{\dag}\na^\nu U\xi^{\dag}
   +\xi^{\dag}\na^\mu U\na^\nu U^{\dag}\xi)>.
\end{eqnarray}
Therefore, every tadpole-loop contributes a factor ($-2\zeta$) to
$\rho\rightarrow\vphi\vphi$ vertex. To sum over all potential tadpole
diagram correction, we obtain
\begin{equation}\label{4.34}
{\cal L}^{(1)}_{\rho\vphi\vphi}=\frac{i}{4}\frac{b(p^2)}{1+2\zeta}
  <\rho_{\mu\nu}[\pa^\mu\vphi,\pa^\nu\vphi]>.
\end{equation}
where $\vphi$ field has been normalized.

\begin{description}
\item[b.] {\bf Two-point corrector}
\end{description}

To replace $b(p^2)/(1+2\zeta)$ in Eq.(~\ref{4.34}) by $b_\gamma(p^2)$ in
Eq.(~\ref{4.23}), we can see that the calculation on two-point corrector
of $\rho\rightarrow\vphi\vphi$ vertex will be the same as one in section
4.2.2. The result can be obtained from Eq.(~\ref{4.29}) directly,
\begin{equation}\label{4.35}
G^{\rho\pi\pi}=\frac{i}{4}\frac{b(p^2)}{1+2\zeta}
  \Sigma(p^2)<\rho_{\mu\nu}[\pa^\mu\pi,\pa^\nu\pi]>.
\end{equation}
To sum over all diagrams of chain approxiamtion in figure(4.4), we
obtain the complete $\rho\rightarrow\pi\pi$ vertex,
\begin{equation}\label{4.36}
{\cal L}_{\rho\pi\pi}^{c}=-\frac{i}{4}g_{\rho\pi\pi}(p^2)
  <\rho_{\mu\nu}[\pa^\mu\pi,\pa^\nu\pi]>,
\end{equation}
where
\begin{equation}\label{4.37}
g_{\rho\pi\pi}(p^2)=\frac{b(p^2)}{(1+2\zeta)(1+\Sigma(p^2))}.
\end{equation}

\subsection{Correction to $\rho-\gamma$ vertex}

The complete one-loop correction to $\rho-\gamma$ vertex contains two
different ingrendients:

1) The effective lagrangian(~\ref{3.12}) will generate
tadpole diagram. For obtaining complete tadpole-loop correction, we need
to sum over all tadpole-loop diagrams in figure(4.5).

\begin{figure}[hpt]
\label{chain2}
\centerline{\psfig{figure=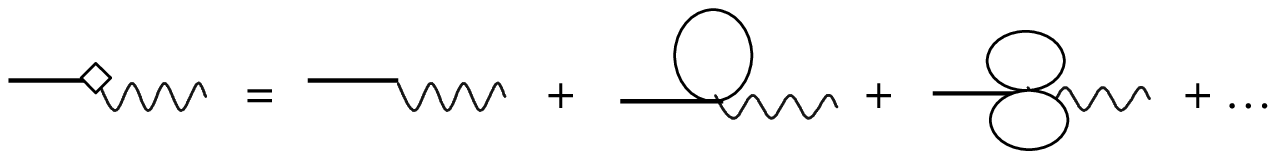,width=5.5in}}
\centering
\begin{minipage}{5in}
   \caption{The $\rho-\gamma$ vertex with correction by all tadpole
   diagrams.}
\end{minipage}
\end{figure}

2) The chain approximation correction in figure(4.6). Here these loop
graphs are generated by $\rho\rightarrow\vphi\vphi$, 4-pseudoscalar and
$\gamma\rightarrow\vphi\vphi$ vertices. All vertices should include
correction of all potential tadpole diagrams. These corrections have been
obtained in section 4.1, 4.2 and 4.3.

\begin{figure}[hpb]
\label{m2}
\centerline{\psfig{figure=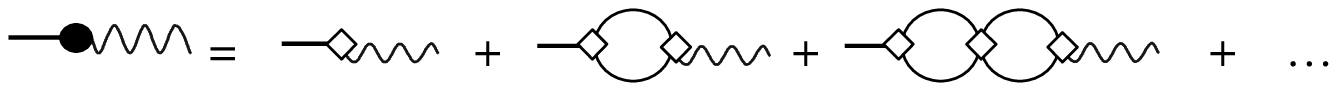,width=5.5in}}
 \centering
\begin{minipage}{5in}
   \caption{Chain approximation to $\rho-\gamma$ couping, which is
    generated by $\rho\rightarrow\vphi\vphi$, 4-pseudoscalar and
    $\gamma\rightarrow\vphi\vphi$ vertices. Here $``\diamond''$ denotes
    every vertices are with complete tadpole-loop correction.}
\end{minipage}
\end{figure}

Note that Only pseudoscalar mesons in SU(3)/SU(2) sector yield
tadpole-loop contribution. Then fig.(4.5) and fig.(4.6) lead to complete 
$\rho-\gamma$ coupling vertex as follow
\begin{equation}\label{4.38}
{\cal L}_{\rho\gamma}^{c}=-\frac{b_{\rho\gamma}(p^2)}{4}
  <\rho_{\mu\nu}\gamma^{\mu\nu}>,
\end{equation}
where
\begin{equation}\label{4.39}
b_{\rho\gamma}(p^2)=\frac{A(p^2)}{g(1+\zeta)}
   -f_\pi^2b(p^2)\frac{\Sigma_0(p^2)}{1+2\zeta}
  [1+\frac{p^2b_{\gamma}(p^2)}{1+\Sigma(p^2)}].
\end{equation}

\subsection{Unitarity and propagator of $\rho$-meson}

The unitarity must be staified for every reliable theory. In this
subsection we will examine unitarity of this present formalism via forward
scattering of $\rho$-meson. The examination on other processes can be
performed similarly. We define $S$-matrix and ${\cal T}$-matrix as usual,
\begin{equation}\label{4.40}
<\beta|S(=Te^{i\int d^4x{\cal L}(x)})|\alpha>
  =S_{\beta,\alpha}
  =\delta_{\beta,\alpha}+i\delta^{(4)}(p_\beta-p_\alpha){\cal
  T}_{\beta,\alpha}.
\end{equation}
The unitarity requires
\begin{equation}\label{4.41}
{\rm Im}{\cal T}_{\beta,\alpha}=\frac{1}{2}\int d\Psi \delta^{(4)}
  (p_{\Psi}-p_{\alpha}){\cal T}_{\Psi,\alpha}^{*}
  {\cal T}_{\Psi,\beta},
\end{equation}
where $\Psi$ is all potential physics states. For the case of
$\alpha=\beta=\rho$, $<\Psi|=<\pi\pi|$ is dominant. Then for
forward scattering of $\rho$-meson, Eq.(~\ref{4.31}) becomes
\begin{equation}\label{4.42}
\Gamma(\rho\rightarrow\pi\pi)=\frac{2}{(2\pi)^4}{\rm Im}{\cal
   T}_{\rho\rho}.
\end{equation}
The width $\Gamma(\rho\rightarrow\pi\pi)$ can be obtained from the
complete vertex(~\ref{4.36}),
\begin{equation}\label{4.43}
\Gamma(\rho\rightarrow\pi\pi)=\frac{|g_{\rho\pi\pi}(m_\rho^2)|^2m_\rho^5}
  {48\pi}(1-\frac{4m_\pi^2}{m_\rho^2})^{3/2}.
\end{equation}

For obtaining ${\rm Im}{\cal T}_{\rho\rho}$, we need to calculate
chain-approximation correction of pseudoscalar loops for two-point vertex
of $\rho$-meson. The calculation is similar to one in section 4.3. To use
the equation of motion of $\rho$-meson, eq.~(\ref{3.9}), and renormalize
the mass of $\rho$-meson, we have
\begin{equation}\label{4.44}
{\cal L}_{\rho\rho}^{1-{\rm loop}}=i{\rm Im}[\Sigma_0(m_\rho^2)
    \frac{f_\pi^2m_\rho^4b^2(m_\rho^2)}
    {g^2(1+2\zeta)^2(1+\Sigma(m_\rho^2))}]\rho_\mu^i\rho^{i\mu}.
\end{equation}
where $\rho$-field has been normalized.

Since Im$(\Sigma_0\cdot\Sigma)\equiv 0$, we obtain
\begin{equation}\label{4.45}
\frac{2}{(2\pi)^4}{\rm Im}{\cal T}_{\rho\rho}=-2[{\rm
  Im}\Sigma_0(m_\rho^2)]
 \frac{f_\pi^2m_\rho^3g^{-2}b^2(m_\rho^2)}
 {(1+2\zeta)^2|1+\Sigma(m_\rho^2)|^2}.
\end{equation}
Setting $Arg(-1)=-\pi$ in Eq.(~\ref{4.48}), we have
\begin{eqnarray}\label{4.46}
{\rm Im}\Sigma_0(m_\rho^2)=-\frac{m_\rho^2}{24\pi f_\pi^2}
  \theta(m_\rho^2-4m_\pi^2)\equiv-\frac{m_\rho^2}{24\pi f_\pi^2}. 
\end{eqnarray}
Inserting Eq.(~\ref{4.46}) into Eq.(~\ref{4.45}) and comparing with
Eq.(~\ref{4.43}), we can see unitary condition (\ref{4.42}) is satisfied
at the limit of massless pion. In addition, the difference between
Eqs.(~\ref{4.43}) and (\ref{4.46}) implies that we can perform the
following replacement  
\begin{eqnarray}\label{4.47}
 \theta(m_\rho^2-4m_\pi^2)\rightarrow (1-\frac{4m_\pi^2}{m_\rho^2})^{3/2}.
\end{eqnarray}
This replacement will compensate for the approximation of massless pion.
In terms of similar method, we can also prove unitarity on
$\rho^0\rightarrow\gamma\rightarrow e^+e^-$ decay.

\begin{figure}[hptb]
\centerline{\psfig{figure=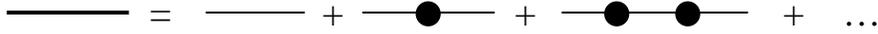,width=5in}}
\centering
\begin{minipage}{5in}
   \caption{The complete propagator of $\rho$-meson. Here
   ``$\bullet$'' denotes the two-point vertex generated by meson-loop.}
\end{minipage}
\end{figure}

Finally, the complete propagator of $\rho$-meson can be obtained via
chain approximation in figure(4.7), where every ``$\bullet$'' denotes a
two-point vertex (~\ref{4.44}). The result is
\begin{equation}\label{4.48}
\Delta_{\mu\nu}^{(\rho)}(p^2)=\frac{-i}{p^2-m_\rho^2
  +im_\rho\Gamma_\rho}(\delta_{\mu\nu}-
  (\propto \frac{p_\mu p_\nu}{m_\rho^2})\;{\rm term}).
\end{equation}
where the width $\Gamma_\rho\simeq \Gamma(\rho\rightarrow\pi\pi)$ which is
given in Eq.(~\ref{4.43}). In this paper since we treat $\rho$-meson at
tree level, the $(\propto p_\mu p_\nu/m_\rho^2)$ term in the
propagator is unimportant. Then the propagator(~\ref{4.48}) is just
well-know Breit-Wigner formula for resonances.

\subsection{Cancellation of divergence}

From the above calculations we can find that there is only quadratic
divergence appears in one-loop contribution of pseudoscalar mesons.
Since the present model is a non-renormalizable effective theory, these
divergences have to be factorized, i.e., the parameter $\lambda$ has to
be determined phenomenologically. Fortunately, this parameter can be
fitted by Zweig rule.

The on-shell decay $\phi\rightarrow\pi\pi$ is forbidden by G parity
conservation and Zweig rule. Experiment also show that branching ratios of
this decay is very small, $B(\phi\rightarrow\pi\pi)=(8\;{+5\atop
-4})\times 10^{-5}$. Theoretically, this decay can occur through
photon-exchange or $K$-loop(figure(4.8)). The latter two diagrams
yield non-zero imagnary part of decay amplitude. Thus the real part
yielded by the latter two diagrams should be very small. We can determine
$\lambda$ due to this requirement(the chiral expansion in powers of
$\phi$-mass will be studied in other papers, but it do not affect
us to fit $\lambda$ here). From the calculation in the above section, we
see that the result yielded by the latter two diagrams is proportional to
a factor
\begin{equation}\label{4.49}
\lambda(\frac{p^2}{2}-m_{_K}^2)-\int_0^1dx\cdot [m_{_K}^2-x(1-x)p^2]
   \ln{(1-\frac{x(1-x)p^2}{m_{_K}^2})}|_{p^2=m_{\phi}^2}.
\end{equation}
Since we only focus our attention on real part of the above equation,
$m_{\phi}^2=4m_{_K}^2$ is a enough approximation. Then Zweig rule requires
that
\begin{equation}\label{4.50}
\lambda(\frac{p^2}{2}-m_{_K}^2)-Re\{-\int_0^1dx\cdot
 [m_{_K}^2-x(1-x)p^2]\ln{(1-\frac{x(1-x)p^2}{m_{_K}^2})}\}
 |_{p^2=4m_{_K}^2}\simeq 0.
\end{equation}
Form the above equation, we obtain
\begin{equation}\label{4.51}
\lambda\simeq\frac{2}{3}.
\end{equation}

\begin{figure}[hptb]
\centerline{\psfig{figure=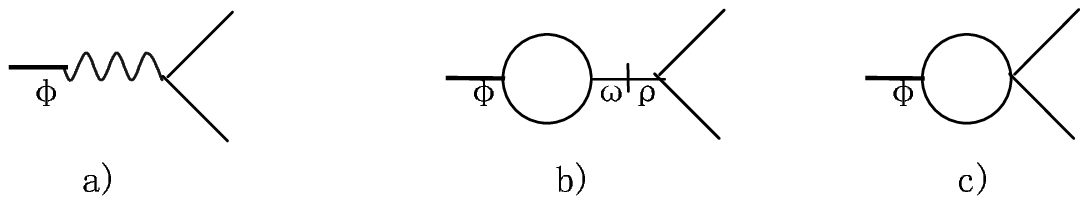,width=5in}}
\centering
\begin{minipage}{5in}
   \caption{Some diagrams for $\phi\rightarrow\pi\pi$ decay. The   
   one-loop in figure b) and c) is $K$-loop.}
\end{minipage}
\end{figure}

\section{Numerical Results and Discussion}
\setcounter{equation}{0}
\setcounter{figure}{0}

In this section we will calculate on-shell $\rho\rightarrow\pi\pi$ and
$\rho^0\rightarrow e^+e^-$ decay numerically. The following parameters
will relate to our calculation: The constituent quark mass $m\simeq
480$MeV is fitted by chiral coupling constants of ChPT at $p^4$. The
universal coupling constant $g=\pi^{-1}$ is determined by KSRF(I) sum rule
and $\lambda\simeq \frac{2}{3}$ is determined by Zweig rule. Other
parameters $f_\pi=185$MeV, $m_\rho=770$MeV and $m_{_K}=495$MeV are fitted
by experimental data. The formula for $\rho\rightarrow\pi\pi$ decay width
has been given in Eq.(~\ref{4.43}), and the width of $\rho^0\rightarrow
e^+e^-$ decay is given as follow
\begin{equation}\label{5.1}
\Gamma(\rho^0\rightarrow e^+e^-)=\frac{4\pi}{3m_\rho^3}
  |g_{\rho\gamma}(m_\rho^2)|^2\alpha^2,
\end{equation}
with
\begin{equation}\label{5.2}
g_{\rho\gamma}(p^2)=\frac{b_{\rho\gamma}(p^2)}{2g}p^2,
\end{equation}
where $b_{\rho\gamma}$ was given in Eq.(~\ref{4.39}). Then we obtain
$\Gamma(\rho\rightarrow\pi\pi)$=146MeV and $\Gamma(\rho^0\rightarrow
e^+e^-)$=7.0MeV. These value agree with data excellently.

\begin{table}[pb]
\centering
\tabcolsep 0.25in
 \begin{tabular}{c|cccc}
            &  I) & II) & III) & Expertiment \\ \hline
  $\Gamma(\rho\rightarrow\pi\pi)$(MeV)& 125& 182& 146 &150\\
  $\Gamma(\rho^0\rightarrow e^+e^-)$(KeV)&4.32 & 7.76 & 7.0 &
   $6.77\pm 0.32$ \\
  $|f_{\rho\pi\pi}(m_\rho^2)|$& 5.51 & 6.62 & 5.94 & - \\
  $|g_{\rho\gamma}(m_\rho^2)|$(GeV$^2$)&0.094& 0.126 &0.12 & -\\   
  $\frac{1}{2}|f_{\rho\pi\pi}(m_\rho^2)|f_\pi^2$(GeV$^2$)&
  0.094 &$0.114^{a)}$&$0.1^{a)}$ & -\\
   \end{tabular}
\begin{minipage}{5in}
\caption {\small The widths of $\rho\rightarrow\pi\pi$ and
$\rho\rightarrow e^+e^-$ decays. I) The values at leading order of
momentum expansion and without meson-loop correction. II) The values with
all order of momentum expansion but without meson-loop correction. III)
The values with all order momentum expansion and with meson-loop
correction. a) Comparing $\frac{1}{2}f_{\rho\pi\pi}(m_\rho^2)f_\pi^2$ with
$g_{\rho\gamma}(m_\rho^2)$, we can obtain the broken of KSRF(I) sum
rule.}
 \end{minipage}  
\end{table}

In table 2 we compare the widths of $\rho\rightarrow\pi\pi$ and
$\rho\rightarrow e^+e^-$ decays for three different cases. It clearly
shows that, both of the high order terms of the momentum expansion and
pseudoscalar meson-loop play very important role in chiral expansion at
$m_\rho$-scale. The reason is obvious, that at this energy scale,
the momentum expansion converge slowly, and it is not enough to
merely consider the leading order terms of $N_c^{-1}$ expansion(or
meson-loop expansion). Thus in a reliable and consistent field
theory describing physics at vector meson energy scale, the leading
order theoretical prediction must not agree with experimental data.
Otherwise the important high order correction will become
incomprehensible in logic and phenomenology. For example, from
table 2 we can see that the correction of meson-loop is about
$20\%$. It is agree with $N_c^{-1}$ expansion for $N_c=3$ very
well. In particular, we can not understand the unitarity of this
model at all if our studies are limited to capture merely the
leading order effects of large $N_c$ expansion.

In table 2 we also show how KSRF(I) sum rule is broken. We can see
that both of high power terms of momentum expansion and meson-loop break
KSRF(I) sum rule. This mechanism is agree with experiment very well.

\section{Summary}

The physics on vector meson resonances has been studied continually by
various chiral models during the last two decades. It is well known,
however, that all past studies on this sort of chrial models suffer two
difficulties: 1) The convergence of the chiral expansiom in the models
is unclear. 2) There is no well-defined way to calculate the next to
leading order. This makes the model's calculations being not controlled
approximations in that there is no well-defined way to put error bars on
the predictions. In this present paper, we have provided a self-consistent
pattern to overcome the difficuties mentioned above, that is the ChCQM
formalism.
The chiral constituent quark model with vector meson fields is
formulated only by two basic ideas: One is transformation properties of
relevant fields under SU(3)$_{_V}$ and another one is to treat vector
mesons as composited fields of constituent quarks. Employing ChCQM, we
have provided a systematical method to investigate the chiral expansion
up to all order, and to perform the calculation to the next leading order
of $N_c^{-1}$ expansion. The results are factorized in $f_\pi$,
$m_0(m_\rho)$, $B_0$, $g_A(=0.75$, $\beta$ decay of neutron),
$g(=\pi^{-1}$, KSRF(I) sum rule), $m(=480$MeV, chiral coupling constants
at $p^4$) and $\lambda(=2/3$, Zweig rule). There are no adjustable
parameters in the theoretical calculations presented in this formalism. By
using this method, $\rho\rightarrow\pi\pi$ and $\rho\rightarrow e^+e^-$
decays are calculated. The predictions are in quite well agreement with
the data. 

Consequently, we conclude that, in ChCQM formalism we can derive a
self-consistent effective field theory with the lowest vector meson
resonances, and the calculation pattern presented in this paper is
legitimat.  

The investigation of this paper reveals the following important features
of effective field theory with the lowest vector meson resonances:

i) The chiral expansion at this energy scale is convergent. The
convergence of chiral expansion is the most important criterion to examine
whether a chiral model including meson resonances can construct a
consistent effective field theory. From this point, many approachs can
only be thought of phenomenological models available at the leading order
of the chiral expansion, since in those models it is diffcult to yield
convergent chiral expansion at vector mesom energy scale.

ii) The chiral expansion at this energy scale converge slowly.
Theoretically, it has been shown by agurement in ChPT, that the chiral
expansion at energy scale $\mu$ should be in powers of
$\mu^2/\Lambda^2_{\rm CSSB}$. Therefore, at $\mu\sim m_\rho$,
complete theoretical predictions have to include high order terms of
the chiral expansion, and the method of ChPT becomes impratical. Thus in
this paper, we studied the chiral expansion in powers of $m_\rho$
systematically by means of the approach of the chiral constituent quark
model. The advantage of this approach is that we can study the chiral
expansion up to all orders but without extra free parameters. Although the
number of parameters is even less than $O(p^4)$ ChPT, this theory's
prediction potential is quite powerful.

iii) The large $N_c$ expansion argues that both of width of meson
resonances and loop effects of mesons are suppressed\cite{tH74}.
For example, in those processes relating to $\rho$-resonance, the
contribution from meson loops is about $\Gamma_{\rho}/m_{\rho}\sim
20\%$. This arguement is comfirmed by unitarity of the chiral
theory(e.g., see Eq.(~\ref{4.42})). Thus the loop effects of mesons
also play important role in a chiral effective theory. In this
paper, we study one-loop effects of pseudoscalar mesons systematically.
The logarithmic divergence and quadratic divergence from meson
loops are cancelled by $O(p^4)$ coupling constants of ChPT and
Zweig rule respectively. The contribution from meson loops is about
$20\%-30\%$, which agree with large $N_c$ arguement very well. More
important, unitarity of this chiral effective theory is examined
explicitly, and the one-loop correction of pseudoscalar mesons make
theoretical predition close to experiment. It shows that
calculation on one-loop graphs of mesons is self-consistent. All of
these imply that precise prediction of a chiral meson effective
theory must include the contribution from meson loops.

iv) The low energy limit of this model agree with ChPT very well.
It means that at very low energy, this effective model will return
to ChPT.

v) Those phenomenological successful ideas, such as VMD and universal
coupling for vector meson resonances, can be predicted by this effective
theory. It is quite nature in the formalism of ChCQM.

Finally, the calculation on $\rho$-meson in this paper can be easily
extend to cases including $K^*(892)$ and $\phi(1020)$. The difference is
that the strange quark mass play important role in the chiral expansion at
$m_{_{K^*}}$ or $m_{\phi}$-scale. These studies will be found elsewhere.     

\begin{center}
{\bf ACKNOWLEDGMENTS}
\end{center}
We would like to thank Prof. D.N. Gao and Dr. J.J. Zhu for their helpful
discussion. This work is partially supported by NSF of China through C. N.
Yang and the Grant LWTZ-1298 of Chinese Academy of Science.

\end{document}